\documentclass[preprint]{elsarticle} 
\usepackage{ucs}
\usepackage[utf8x]{inputenc}
\usepackage{amsmath,amssymb,amsfonts,mathtools}
\usepackage[english]{babel}
\usepackage{fontenc}
\usepackage{graphicx}
\usepackage{ulem} 
\usepackage[margin=0pt,font=footnotesize, labelsep=colon]{caption}
\usepackage{tabularx}
\usepackage{multirow}
\usepackage{siunitx}
\usepackage{booktabs}
\usepackage{color}
\usepackage[table]{xcolor}
\usepackage{colortbl}
\usepackage{supertabular}
\usepackage{array}
\usepackage{soul}
\usepackage{subcaption}

\sethlcolor{green}
\definecolor{rgrey}{gray}{0.75}

\usepackage{tikz}
\newcommand*\circled[1]{\tikz[baseline=(char.base)]{
            \node[shape=circle,draw,inner sep=2pt] (char) {#1};}}

\DeclareMathOperator{\Ker}{Ker}

\begin{document}

\begin{frontmatter}

\title{Validation of Danish wind time series from a new global renewable energy atlas for energy system analysis}

\author[label1]{Gorm B. Andresen}
\ead{gba@eng.au.dk}
\author[label1,label2]{Anders A. S\o{}ndergaard}
\author[label1,label3]{Martin Greiner}
\address[label1]{Department of Engineering, Aarhus University, Inge Lehmanns Gade 10, 8000 Aarhus C,  Denmark}
\address[label2]{Department of Physics and Astronomy, Aarhus University, Ny Munkegade 120, 8000 Aarhus C,  Denmark}
\address[label3]{Department of Matematics, Aarhus University, Ny Munkegade 118, 8000 Aarhus C,  Denmark}
  
\date{\today}

\begin{abstract}
We present a new  high-resolution global renewable energy atlas ({REatlas}) that can be used to calculate customised hourly time series of wind and solar PV power generation. In this paper, the atlas is applied to produce 32-year-long hourly model wind power time series for Denmark for each historical and future year between 1980 and 2035. These are calibrated and validated against real production data from the period 2000 to 2010. The high number of years allows us to discuss how the characteristics of Danish wind power generation varies between individual weather years. As an example, the annual energy production is found to vary by $\pm$10\% from the average. Furthermore, we show how the production pattern change as small onshore turbines are gradually replaced by large onshore and offshore turbines. Finally, we compare our wind power time series for 2020 to corresponding data from a handful of Danish energy system models. The aim is to illustrate how current differences in model wind may result in significant differences in technical and economical model predictions. These include up to 15\% differences in installed capacity and 40\% differences in system reserve requirements.
\end{abstract}


\end{frontmatter}
\section{Introduction}
\label{section:Introduction}
In 2020, Danish wind power is expected to cover about 50\% of the Danish electricity consumption on average \cite{Danish-Energy-Agency:2010uo}. This means that fluctuations in the wind will clearly dominate many technical and economical aspects of the power system. For this reason, it is of obvious importance to use an accurate representation of the future wind power in models of the future power system. But to the best of our knowledge, there is currently no consensus on how to select and validate a good forecast of future wind power generation time series for the different energy system models of e.g. Denmark.

In this paper, we present and apply a new global high-resolution renewable energy atlas ({REatlas}) to model hourly Danish wind power generation for all years between 1980 and 2035. The model is based on a detailed representation of historical and future configurations of Danish wind turbines. With a simple calibration of the wind speed-to-power conversion, we show how historical wind power production for Denmark during the period 2000 to 2010 can be reproduced. The calibrated model is then used to convert 32 years of weather data to hourly model time series for turbine configurations representing each year from 1980 to 2035. These are available for download in the Python npy-format \cite{Andresen:2013uq}. To the best of our knowledge, this study represents the most detailed model calculations of future Danish wind power generation available in the scientific literature.

The time series are used to forecast future characteristics of Danish wind power generation, and to show how past and future wind power generation from a particular turbine configuration varies between individual weather years. In addition, we compare and discuss wind power time series for 2020 from different Danish energy system analysis tools to illustrate how large and important differences in the prediction of technical and economical characteristics of the future power system can be caused by differences in the wind power time series. We are not aware that such a comparison has been carried out previously.\\

The {REatlas} used in this study is global and can be applied to generate hourly time series of both wind and solar {PV} power generation. It is based on a state-of-the-art global 32-year-long meteorological data set from the American weather service NOAA, and it meets a number of design goals for state-of-the-art and next generation energy system modelling (see Section~\ref{section:REatlasDesignGoals}). In this paper, we present the detailed implementation of the {REatlas} for the first time.

The development of the {REatlas} is motivated by a need for an efficient and configurable wind and solar PV conversion tool. As an example, it has been optimised in software and hardware for very fast repeated wind or solar PV conversions with different technologies and/or geographical capacity assignments. Among other things, this makes it possible to make predictions of the consequences of development in wind or solar technologies, different renewable energy strategies, and even optimisation of strategies with respect to multiple objectives. It also allows for easy sensitivity analysis. Other wind or solar atlas with high temporal resolution, e.g. hourly, do not allow the user to modify the input parameters for the conversion to wind or solar power. Instead, specific technologies have been preselected, and only the resulting time series are available (examples are \cite{Bofinger:2008kx,Potter:2009dp,Brower:2009bf}).\\

The paper is organised as follows: Previous, related studies are briefly discussed in Section~\ref{section:PreviousStudies}. In Section~\ref{section:REatlas}, the {REatlas} is described. Section~\ref{section:Validation} is concerned with validation of model wind power time series for Denmark 1980 to 2035. Section~\ref{section:Discussion} contains a discussion of wind power time series for 2020 from a handful of different models of the future Danish power system. Finally, the paper is concluded in Section~\ref{section:Conclusion}. Further details on the {REatlas} software and hardware implementation can be found in \ref{appendix:conversionSetup}.

\section{Previous studies}
\label{section:PreviousStudies}


The time series presented in this paper are based in reanalysis data from state-of-the-art climate data. This approach offer many advantages for the generation of wind power time series. In particular,  multiple decades of consistent global weather records are available in this form, and it contain consistent spatial and temporal correlations. Furthermore, many other weather fields relevant for either weather-driven power generation or demand modelling are included in the climate models. Examples are solar irradiation and temperature, which are both relevant for solar energy production as well as heating or cooling demands.

A number of recent studies use reanalysis data from atmospheric climate models such as ERA-40 \cite{Uppala:2005it} and CFSR \cite{Saha:2010kx} as a source of wind speeds. Most interesting are studies where the resulting wind power time series have been compared against historical records. Kiss et al. \cite{Kiss:2009lq} appears to be first to compared nacelle measurements of wind speed and power generation from two turbines in Hungary to the ERA-40 reanalysis data set. They found a "satisfying agreement" with proper calibration of the conversion model. Other studies include: Hawkins et al. \cite{Hawkins:2011rm} who replicated monthly load factors for the UK from a reanalysis data set calculated with the Weather Research and Forecast (WRF) model. Kubik et al. \cite{Kubik:2013qr} matched the global {NASA} reanalysis to half-hourly power output from wind parks located in Northern Ireland. Finally, Staffel and Green \cite{Staffell:2014xy} used the {NASA} data set to investigated the performance of wind farms located in the UK (United Kingdoms). 

\section{Methodology: {REatlas}}
\label{section:REatlas}

\begin{figure}[tp!]
	\centering	
	\includegraphics[width=\columnwidth]{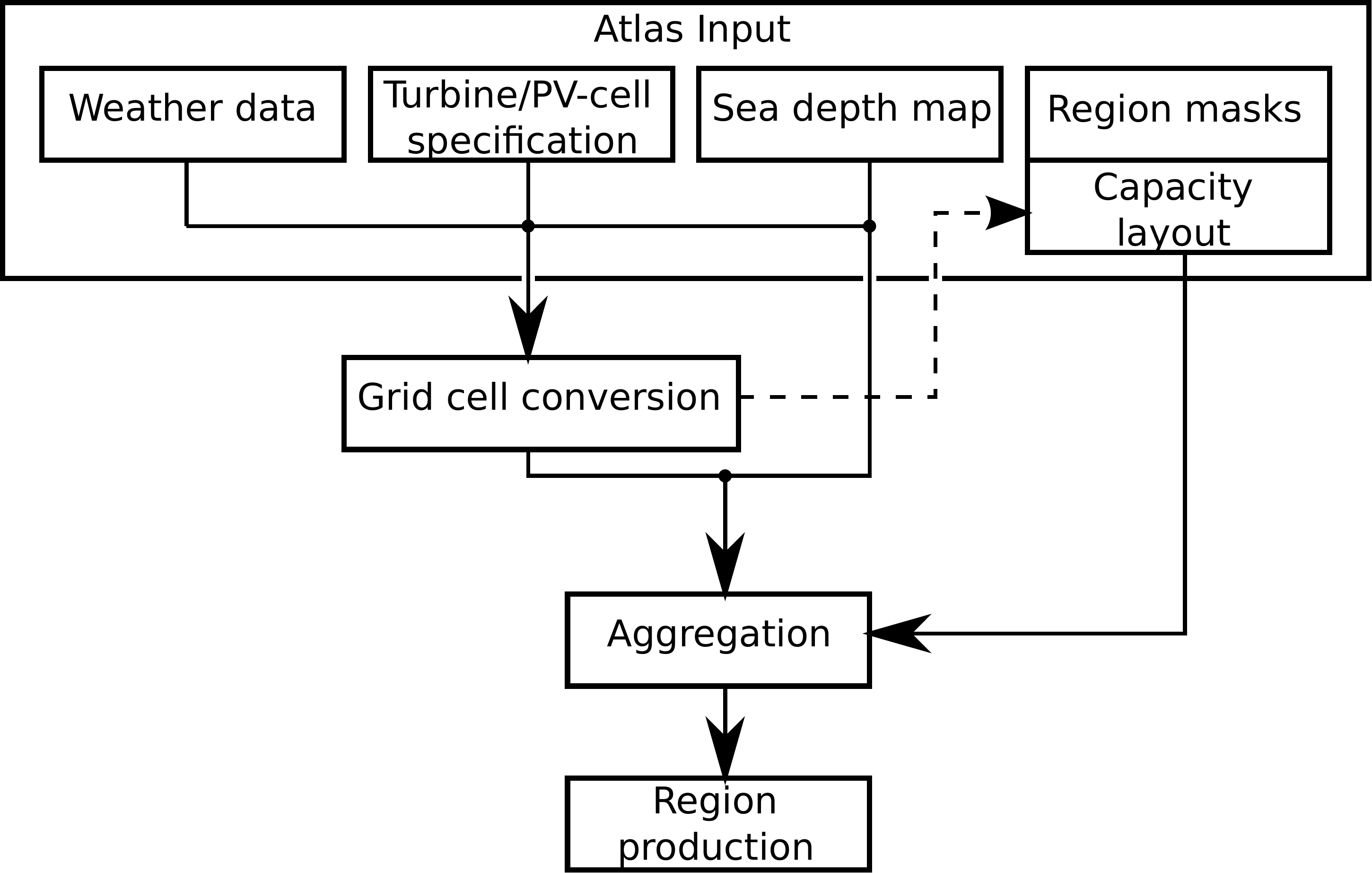}
	
	\caption{Flowchart of the renewable energy atlas. The input data and parameters are first converted to hourly wind or solar PV power generation for each individual grid point. Then the production for a region is obtained by weighted aggregating over the grid cells within a region mask, which specifies the installed capacity for each grid point. A capacity layout proportional to the individual grid point capacity factors can be generated automatically, or a custom layout can be specified by the user.}
	\label{figure:REatlasFlowChart} 
\end{figure}

The {REatlas} is a computer program, currently implemented in a mix between the Python and the C programming languages. It is used to calculate hourly generation profiles of wind and solar PV based on a 32-year-long global high-resolution data set from the American weather service NOAA \cite{Saha:2010kx,Saha:2011od}. Further input is the power curves of wind turbines and the efficiency curves and orientations of solar PV modules. The latter can have fixed orientations or single/dual-axis tracking of the sun. If the total production of a region, e.g. Denmark or Europe is needed, the distribution of installed capacity is also an input, and simple spatial distributions of turbines or PV panels, such as proportionally to generation potential, can be automatically generated. A simplified version of the {REatlas} implementation is shown in Figure~\ref{figure:REatlasFlowChart}.\\

The {REatlas} is designed to satisfy a number of design goals. These are listed in Section~\ref{section:REatlasDesignGoals}. The  input data is described in Sections~\ref{section:weatherData} and \ref{section:SeaDepthData}, and the wind conversion algorithm is described in Section~\ref{section:windConversion}. Details on the solar PV conversion can be found in \cite{Sondergaard:2013fk}. 

Furthermore, a combined software and hardware implementation and optimization of the {REatlas} is described in Section~\ref{section:REatlasImplementation} with additional details in \ref{appendix:conversionSetup}. To be of practical use, the atlas needs to be fast, and because of the relatively large amount of input data, this is a big computational challenge. As a result of the optimization, the conversion time is reduced by a factor of about 10,000 as compared to a first, simple implementation on a standard desktop PC. 

\subsection{Design goals for the {REatlas}}\label{section:REatlasDesignGoals}

\noindent The {REatlas} is designed to satisfy the following list of requirements:\\

\noindent 1. The correlations between wind and solar PV power generation time series are captured both spatially and at all time scales. This is achieved by using weather data from coherent climate model for both wind and solar PV conversions.\\

\noindent 2. The choice of wind or solar PV technology is configurable which allows the user to analyse the impact of different technology. For example, it is possible to study roof mounted solar PV vs. solar farms and the influence of new vs. old wind turbines.\\

\noindent 3. The geographical distribution of installed wind or solar PV capacity is configurable and can be used for, e.g. optimisation and validation purposes (see Section~\ref{section:Validation}).\\

\noindent 4. Wind or solar PV conversions can be calculated relatively quickly even for a large number of data points in time and space (see Section~\ref{section:REatlasImplementation}). This makes it is computationally feasible to iterate over different input configurations to allow for, e.g. optimisation.\\

\noindent 5. Many weather oscillations affect global weather. The most well known is probably the El Ni{\~{n}}o, causing extreme weather such as floods and droughts. These oscillations have period lengths well over a year. To capture the influence of these, 32 years of weather data is currently included in the {REatlas}.\\

\noindent 6. The {REatlas} is based on data from a global climate model, which means that all regions of the world can be studied. In addition, the {REatlas} allows for easy and customisable selection of regions of interest. Inconsistencies near the edge of the region of interest, which are present in some data sets that are based on regional climate models, e.g. \cite{Bofinger:2008kx}, do not occur.\\

\noindent 7. Finally, the underlying weather and load data is easily replaceable with, e.g. new updated data, forecast and ensemble data, or synthetic data.\\

\subsection{Weather data}\label{section:weatherData}

The data used in the {REatlas} is a global 32-year-long high-resolution weather data set called CFSR (Climate Forecast System Reanalysis) from NCEP (American National Centers for Environmental Prediction) \cite{Saha:2010kx}. It is a reanalysis data set, which means that it is made by assimilating observed weather data into a numerical weather prediction model. An advantage of the CFSR data over most other reanalyses is that the same model is used for the entire 32-year-period. This ensures consistency. The weather model is a global, state-of-the-art, coupled atmospheric and oceanic model developed at NCEP called CFS version 2. It is initialised and run with observation data from January 1979 to December 2010.

The model output parameters are freely available in the {GRIB2} file format (essentially a {JPEG} compressed array of floating point numbers). These contain data with a temporal resolution of 1~hour and a spatial resolution of $0.3125^\circ \times 0.3125^\circ$. In Europe, the resolution corresponds to roughly $40\times40$ km$^2$. In total, the data set contains 663,552 grid points for the entire world and 280,512 hours. Storing this amount of data as raw {IEEE} 754 double-precision floating point numbers would take up 1.4~TB of storage for each parameter, e.g. wind speed at 10~m height. The {GRIB2} compression brings this down to about 100~GB, at the cost of long decompression time. On a standard desktop computer of 2012, the decompression time itself is 1 to 2 days per parameter.

Of the many available parameters in the CFSR data set, only those relevant to the wind and solar conversions are included in the {REatlas}. For the wind conversion, described in Section~\ref{section:windConversion}, we use the hourly wind speed at 10~m height and monthly values of surface roughness. The solar conversion (see \cite{Sondergaard:2013fk}) uses hourly upward and downward shortwave radiation flux at surface level and hourly temperature at 2~m height. In total, this amounts to about 700~GB of data stored in the {GRIB2} format.

Recently, additional data has become available from NCEP for 2011 and beyond \cite{Saha:2011od}. This data will be included in the {REatlas} at a later point.

\subsection{Sea depth data}\label{section:SeaDepthData}

\begin{figure}[tp!]
	\centering	
	\includegraphics[width=\columnwidth]{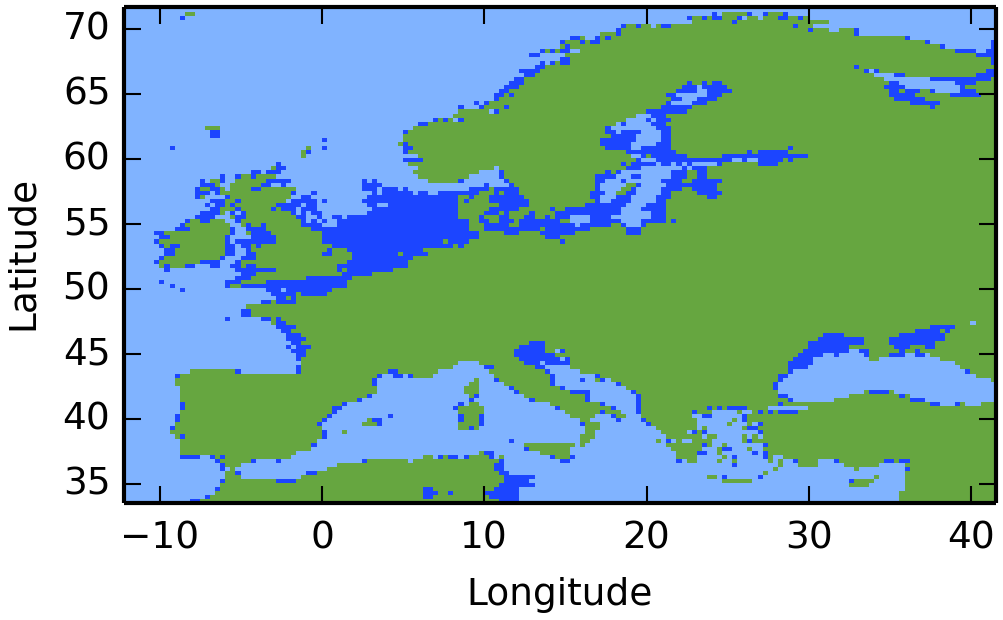}
	
	\caption{Europe (green) with offshore sites where the water depth is less than 70~m (bright blue). It can be seen that Southern Europe has very little area for offshore wind power in shallow water, while the potential is large in the North and Baltic seas. The figure is made with the $0.3125^\circ\times0.3125^\circ$ resolution of the CFSR data set.}
	\label{figure:Europe70mDepth} 
\end{figure}

Modern offshore wind turbines can typically not be installed at water depth below about 70~m. For this reason, a world wide 30~minute resolution bathymetry map, which is freely available from {GEBCO} \cite{GEBCO:2009my}, is included in the {REatlas}. Using a 2D interpolation routine, depth at the centre of each CFSR grid cell is estimated. A map of European offshore grid cells with water depth below 70~m is seen in Figure~\ref{figure:Europe70mDepth}. The resolution of the figure corresponds to the $0.3125^\circ\times0.3125^\circ$ resolution of the {CFSR} data set.

\subsection{Conversion to wind power}\label{section:windConversion}

Wind power generation is approximated from the wind speed $\nu$ at wind turbine hub height $H$ by means of the turbine power curve $P(\nu)$, which is typically available from the manufacturer. Most power curves relate the 10~minute average wind speed at one or more standard turbulence intensities to turbine power output. In the {REatlas}, the user is free to specify any power curve, but the turbulence intensity cannot currently be used to evaluate power generation.

A typical wind turbine power curve is shown in Figure~\ref{figure:PowerCurveModification}. Here, a certain minimum wind speed is required for the turbine to be able to produce useful power. This is called the cut-in speed. The power production then follows a $\nu^3$ behaviour proportional to the kinetic energy flux density in the wind. At rated velocity, the efficiency of the turbine is lowered to limit the power output to the rated power of the turbine. Then, above a certain cut-out wind speed, the turbine is stopped to protect it from high mechanical loads and damage caused by strong wind gusts. For a more detailed description see \cite{Gasch:2004gd}.\\

The wind speed data from the CFSR data set gives the instantaneous wind velocity 10~m above ground/sea, with the zonal and the meridional wind components ($u$ and $v$) interspersed in the {GRIB} file. In the conversion to wind power generation, it is assumed that the turbine is always facing the wind, so only the total speed $\nu$ is used:
\begin{equation}
	\nu = \sqrt{u^2 + v^2}\,.
\end{equation}
Thus, effects related to the wind direction such as wake losses in or between wind farms are not captured.

The wind speed generally increases with height, and a logarithmic wind profile is used to estimate the wind speed at hub height $H$ from the speed at 10~m height \cite{Prandtl:1935tr}:
\begin{equation}\label{equation:windProfile}
	\nu(H) = \nu(10\,\textup{m})\frac{\ln\left({{H}/{z_0}}\right)}{\ln\left({10\,\textup{m}}/{z_0}\right)},
\end{equation}
where $z_0$ is the surface roughness. The actual wind speed profile is more complicated, and among other things, it depends on the stability of the atmosphere, local obstructions such as trees or buildings, and the local orography, i.e. hills and valleys \cite{Oke:1988uo}. The value for surface roughness used here represents an average value for the area around each {CFSR} grid point (about $40\times40$~km$^2$). In complicated onshore terrain, it does not capture the local variations in wind conditions \cite{Badger:2010dp,Badger:2011rf}.

In \cite{Becker:2014jt}, a statistical method described in \cite{Badger:2010dp,Badger:2011rf} was used to modify the wind conversion above  to include local variations in both surface roughness and orography for each grid point in the USA. The effect of selecting only the best sites in each grid cell was studied as well.

\subsection{Optimized software and hardware implementation}\label{section:REatlasImplementation}

The {REatlas} software is primarily written in the high-level language Python with selected subroutines written in C to allow more efficient memory management. The software is currently installed on a Cisco UCS B200 M3 Blade Server equipped with two 2.00~GHz Intel Xenon E5--2650 processors (CPU). Each with 8 hyper threaded cores and 20~MB cache. The machine has a total of 512~GB, 1600~MHz DDR3 memory distributed between four 12.8~GB/s channels per CPU. It is connected to a network storage unit (SAN) with $2\times8$~Gigabit Ethernet.

On this machine, full conversion of all 280,512 hours of a European-wide test region with 21,279~grid points can be performed in about 45~s for wind and less than 2~minutes for solar power. The duration scales approximately linearly with the number of hours and grid points in the subset. It is limited only by a combination of memory bus and CPU processing speeds and not ethernet or hard disk (HD) speeds. When run on a standard 8 threaded desktop PC of 2012, conversion of the same data set takes about 1~h, as it is limited by the HD speed of about 50~MB/s. 

Initial preparation of the test data set takes about 20~h in the current setup. In this step, the global set of weather data from CFSR is decompressed and the subset belonging to the selected region is organised in an uncompressed array. The CFSR data set is compressed one time slice at a time. For this reason, the duration of the initial step scales with the number of hours in the selected subset, but not with the number of grid points. On a standard desktop PC, the initial preparation of a 32-year-long data set takes about a week.

In \ref{appendix:conversionSetup}, the conversion process is described in detail and the choice of hardware is motivated. 

\section{Validation: {REatlas} wind conversion for Denmark}
\label{section:Validation}

The country of Denmark has one of the worlds highest penetrations of wind power in the electricity system, and the Danish wind resource is excellent. This makes Denmark a good test case for the {REatlas}, in particular for the following reasons: i) A detailed model of the distribution of individual turbines can be created because records of the Danish wind turbines are publicly available. These date back to 1980 \cite{Energistyrelsen:2013lq}. ii) The model time series can be compared against historical production data which is available with hourly resolution since 2000 \cite{Energinet.dk:2013xy}. iii) The discrepancy between modelled and real wind power is expected to be low since the terrain is relatively uncomplicated with small variations in height and vegetation (cf. Section~\ref{section:windConversion}).\\

In the following, detailed model representations of the historical and future population of Danish wind turbines are described for all years between 1980 and 2035. For each year, here called a model year, the turbine distribution on January 1$^\textup{st}$ is used as input to the {REatlas} wind conversion to create a model wind power time series for the weather years 1979 to 2010. When the model year overlaps with the weather year, model wind power can be compared directly to historical wind power generation. But due to limited data availability (see ii) above), this is only the case for the period 2000 to 2010. Below, we use data from this period, specifically 2010, to make a simplified calibration of the {REatlas} wind conversion for Denmark. This is required to improve the model performance to compensate for local variations in roughness and orography as well as for possible systematic errors in the underlying weather data. In addition, some variation in the performance and availability of individual wind turbines is to be expected. 

The period 2011 to 2013 does not overlap with the weather years in the {REatlas}, but historical data on the turbine population as well as their historical hourly power generation is available. This allows us to test the calibrated model performance without biasing the calibration. Finally, the calibrated model is used to produce model wind power time series for the future years until 2035, using the expected future turbine populations as input. In particular, we focus on 2020 and 2035.

\subsection{Wind turbine capacities and placement (1980 -- 2035)}\label{section:turbineCapacityAndPlacement}

Maps of the wind turbine capacity layout for all years between 1980 and 2035 are constructed by combining the database of existing turbines ultimo 2013 \cite{Energistyrelsen:2013lq} with the expected annual built-up of new capacity as detailed by the Danish TSO Energinet.dk in  \cite{Energinet.dk:2013fk}. The projections are consistent with current official political targets for the period \cite{DanishGovernment:2011vn}.\\

\begin{figure}[tp!]
	\centering	
	
	\begin{subfigure}[b]{\columnwidth}
		\includegraphics[width=\linewidth]{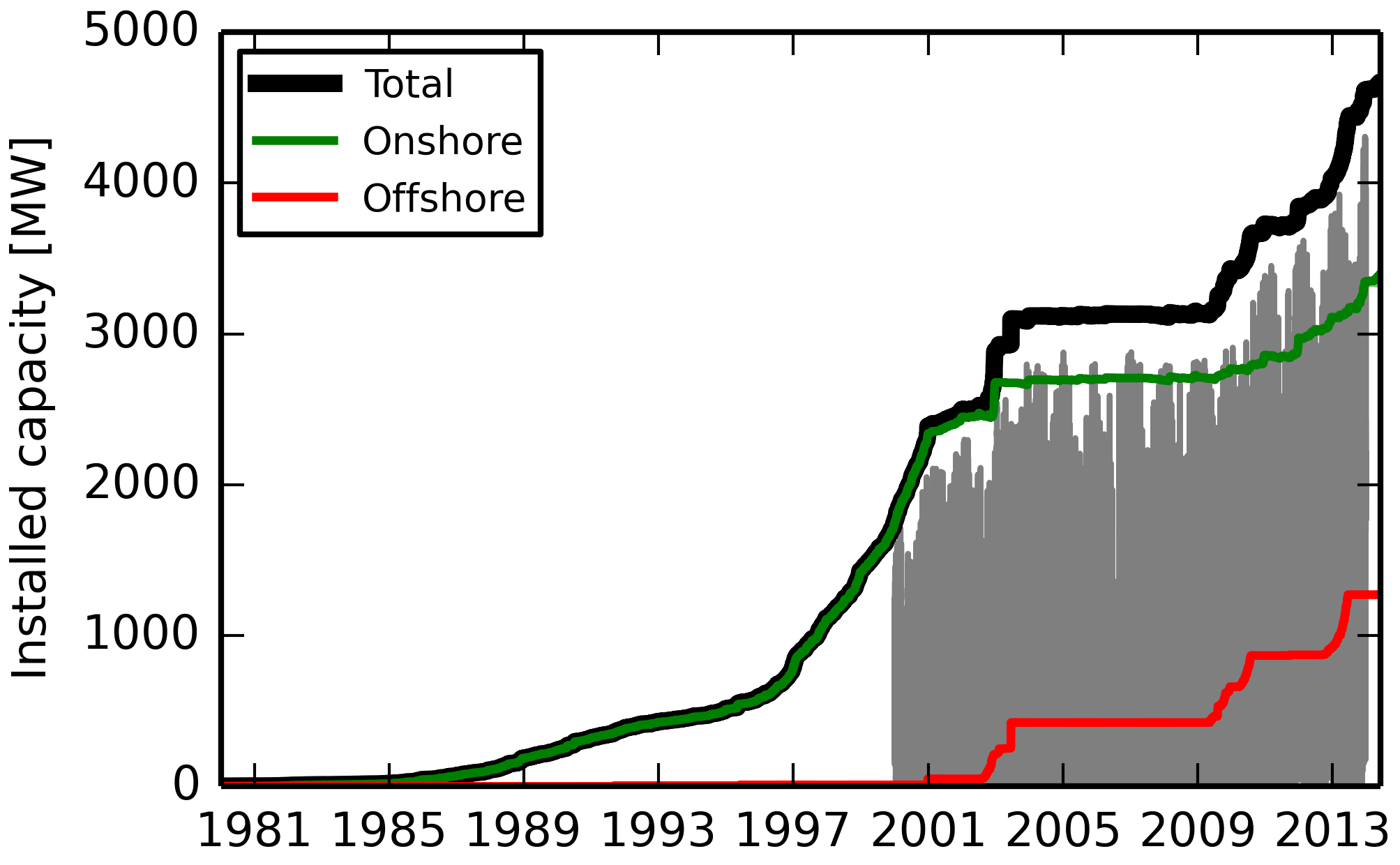}
		\caption{Historical wind power in Denmark.}
	\end{subfigure}
	
	\begin{subfigure}[b]{\columnwidth}	
		\includegraphics[width=\linewidth]{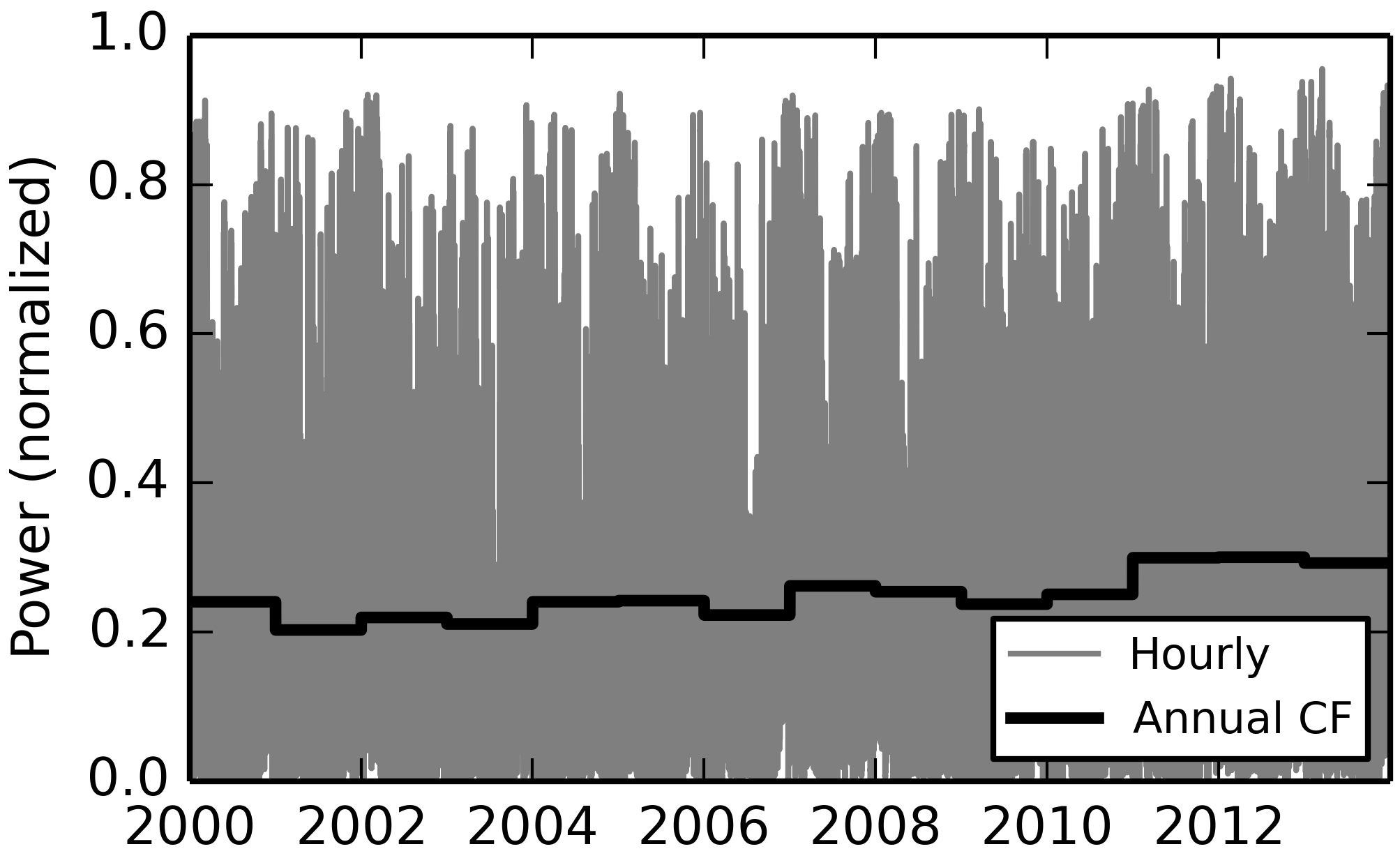}
		\caption{Historical wind capacity factor.}
	\end{subfigure}
	
	\caption{(a) Historical build-up of onshore (green) and offshore (red) wind turbine capacity in Denmark during the period 1980 to 2013. The sum of both onshore and offshore capacity is shown in black. During the period 2000 to 2013, historical production data with hourly resolution is also shown (gray). (b) Hourly power generation for the combined portfolio of Danish wind turbines during the period 2000 to 2013. The black line indicate the annual capacity factor (average of the hourly data). All values are normalised to the total installed capacity on each given day.}
	\label{figure:HistoricalWindCapacities} 
\end{figure}

\begin{figure}[tp!]
	\centering	
	\includegraphics[width=1.0\linewidth]{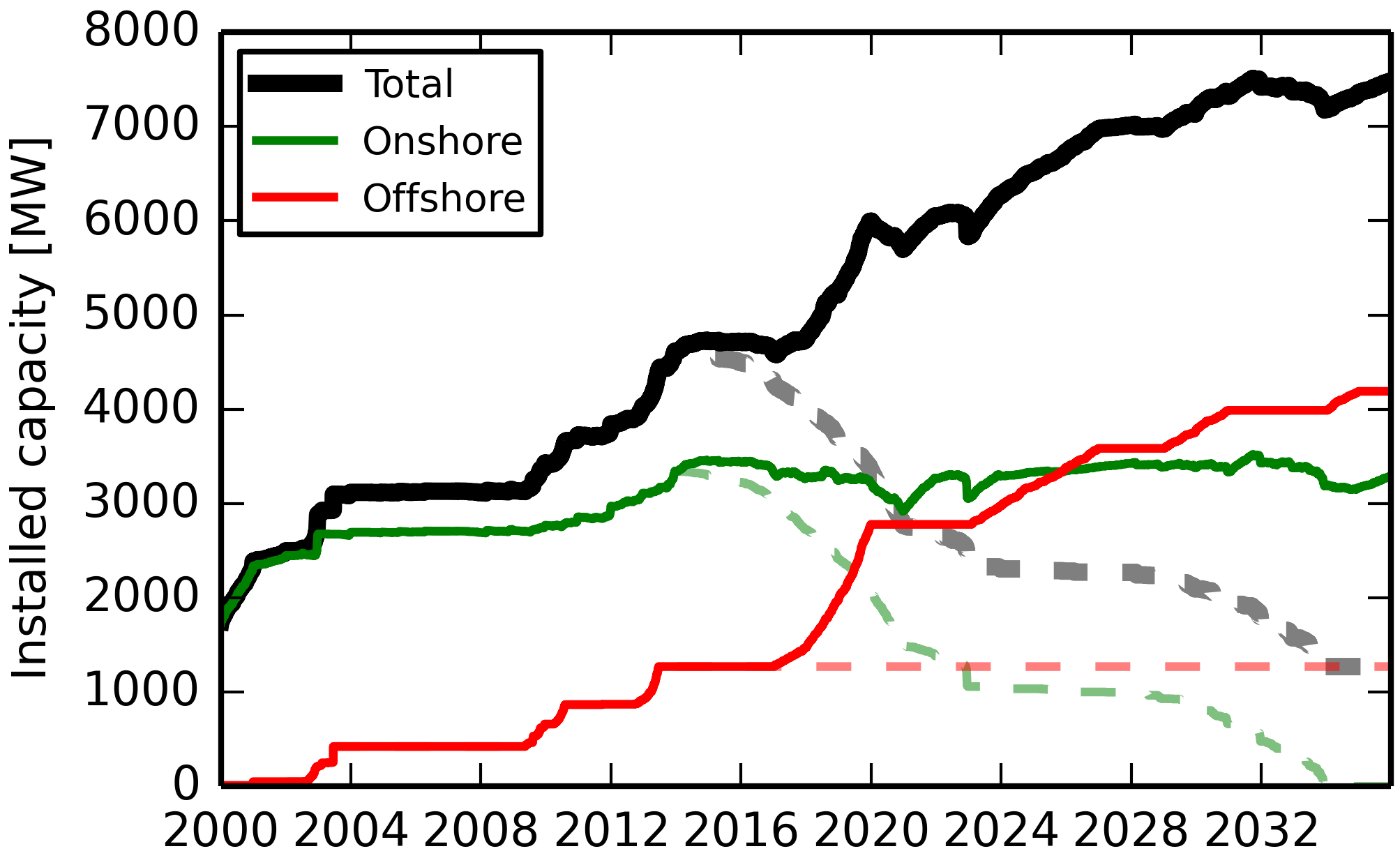}
	\caption{Projected build-up of wind turbine capacity in Denmark during the period 2014 to 2035. Historical data for the period 2000 to 2013 is shown for comparison. Dashed lines indicate what would happen if no new onshore and offshore turbines were build after 2013 (existing offshore capacities are assumed to be maintained until after 2035).}
	\label{figure:ProjectedWindCapacity} 
\end{figure}

All existing onshore turbines are assumed to have a fixed lifetime of 20 years from the day they where connected to the grid. New onshore turbines are also assumed to have a lifetime of 20 years, and their date of grid connection is chosen at random within a year from the year listed in \cite{Energinet.dk:2013fk}. Each new onshore turbine is placed in a random onshore grid cell in one of the two Danish electricity system areas {DK1} or {DK2}, in accordance with \cite{Energinet.dk:2013fk}. It is assigned a name plate capacity of 3,600~MW as this is the current commercial standard. In reality, the capacity could be larger.\\

Offshore turbines are categorized as being either near shore or proper offshore. Here, this distinction is only used when the locations of individual turbines is chosen. Near shore turbines are evenly distributed between five different sites appointed by the Danish government \cite{Energistyrelsen:2012ly} (we exclude the site near Bornholm). All other offshore turbines are located in wind parks with locations as specified in \cite{Energinet.dk:2013fk}. Similar to the onshore turbines, each new offshore turbine is grid connected at a random day within a year from the year listed in \cite{Energinet.dk:2013fk}, and assigned a name plate capacity of 3,600~MW. However, unlike onshore turbines, both existing and future offshore turbines are assumed to stay grid connected until after 2035.\\

The historical build-up of Danish wind turbine capacity is shown in Figure~\ref{figure:HistoricalWindCapacities}, and the expected future build-up is shown in Figure~\ref{figure:ProjectedWindCapacity}. It is evident that the total capacity of onshore turbines is expected to stay at a constant level of about 3,000~MW, with new larger turbines replacing smaller turbines when they are decommissioned. In contrast, about 3,000~MW new offshore capacity will be added to the existing 1,300~MW. Thus, most of the total increase in turbine capacity from about 4,800~MW in 2012 to 7,500~MW in 2035 will be realized as new offshore capacity.\\

\begin{figure*}[tp!]
	\centering	
	
	\begin{subfigure}[b]{\linewidth}
		\includegraphics[width=0.49\linewidth]{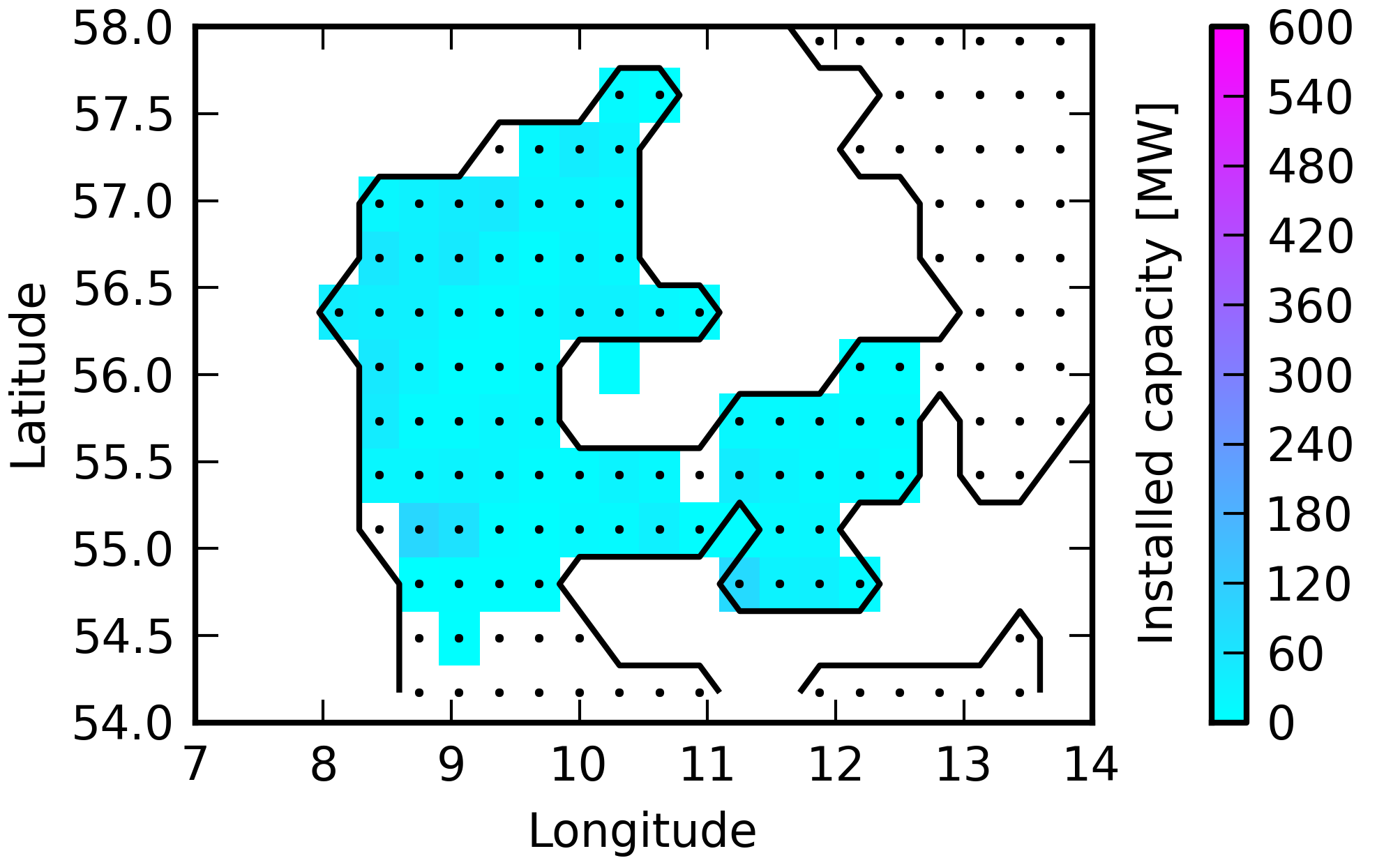}
		\includegraphics[width=0.49\linewidth]{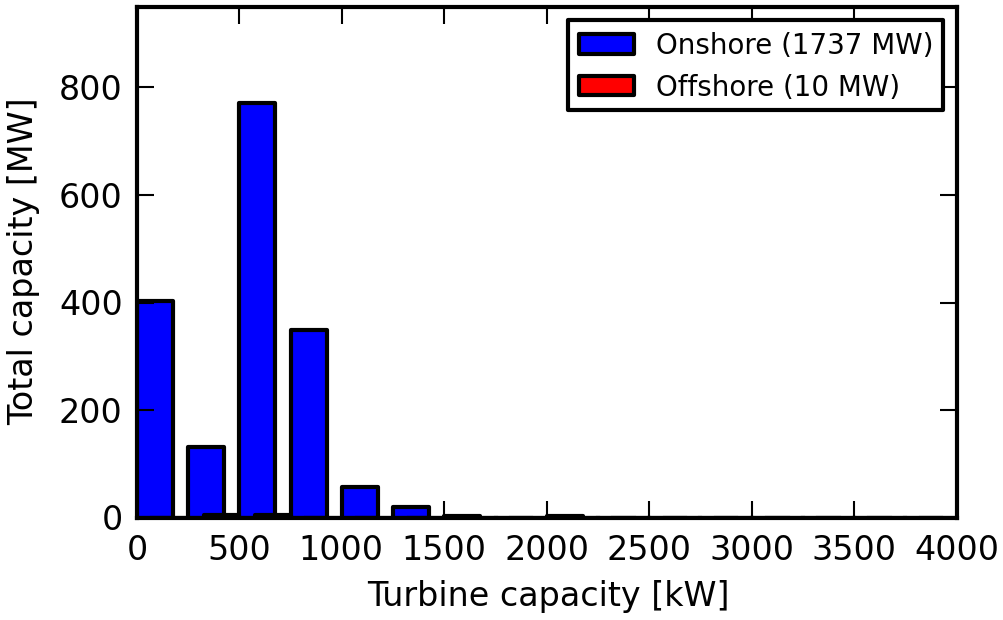}
		\caption{2000: Historical distribution of wind turbines.}
	\end{subfigure}
	
	\begin{subfigure}[b]{\linewidth}
		\includegraphics[width=0.49\linewidth]{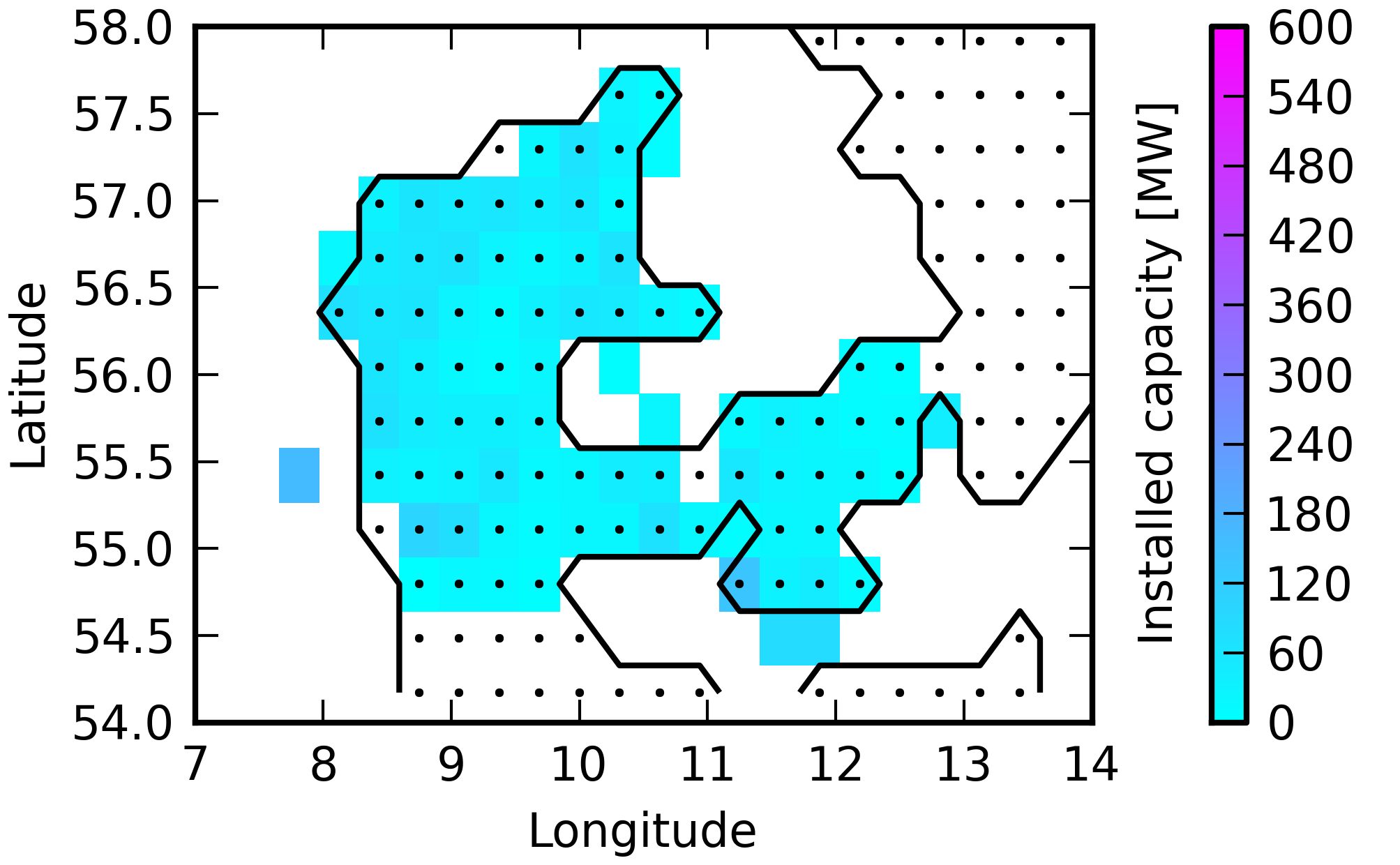}
		\includegraphics[width=0.49\linewidth]{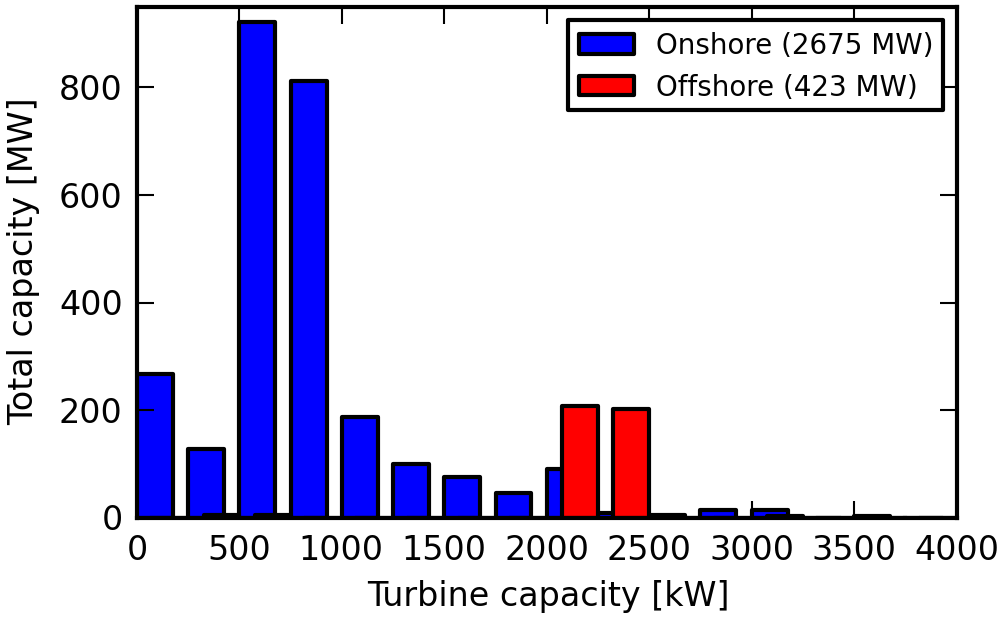}
		\caption{2005: Historical distribution of wind turbines.}
	\end{subfigure}
	
	\begin{subfigure}[b]{\linewidth}
		\includegraphics[width=0.49\linewidth]{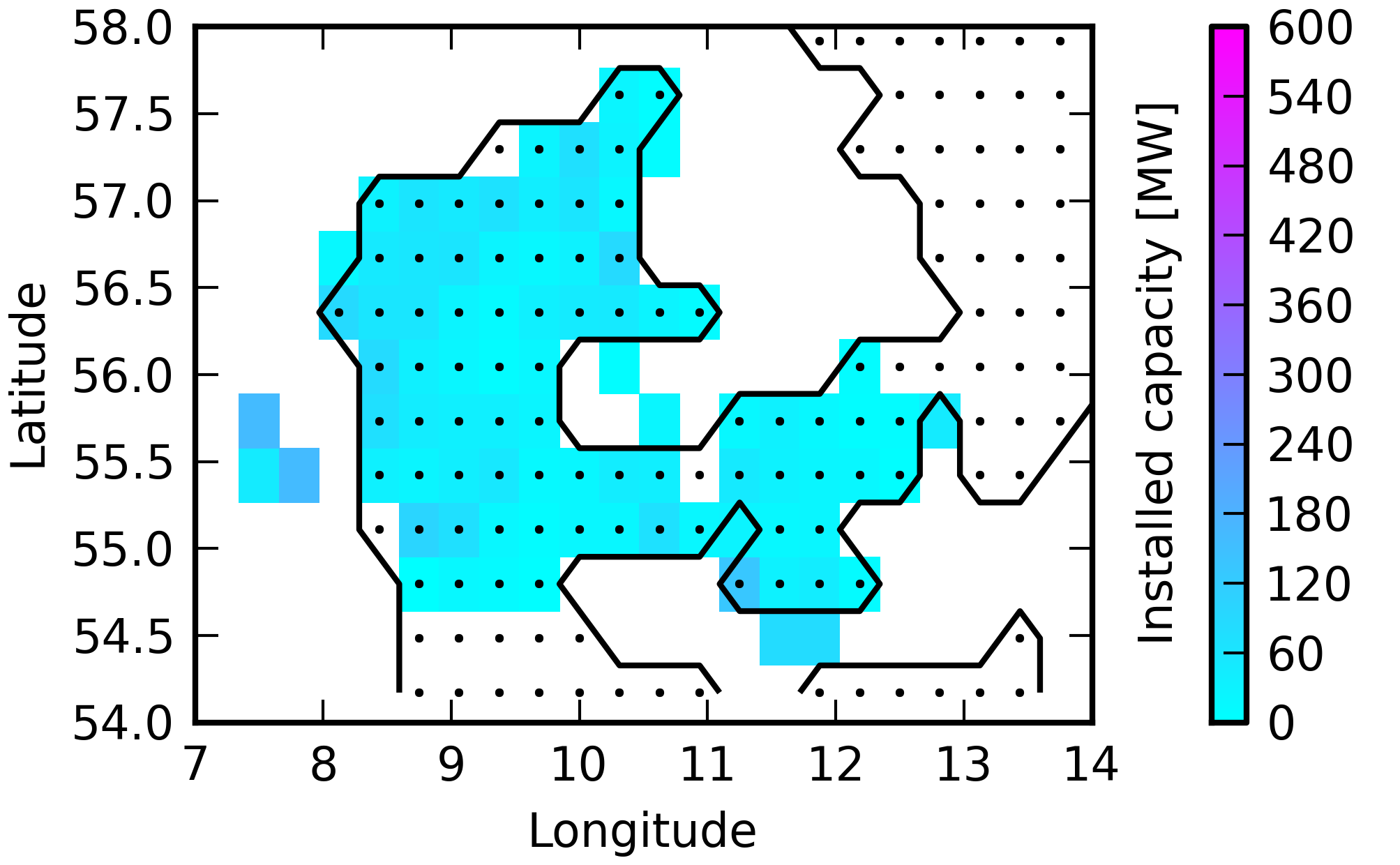}
		\includegraphics[width=0.49\linewidth]{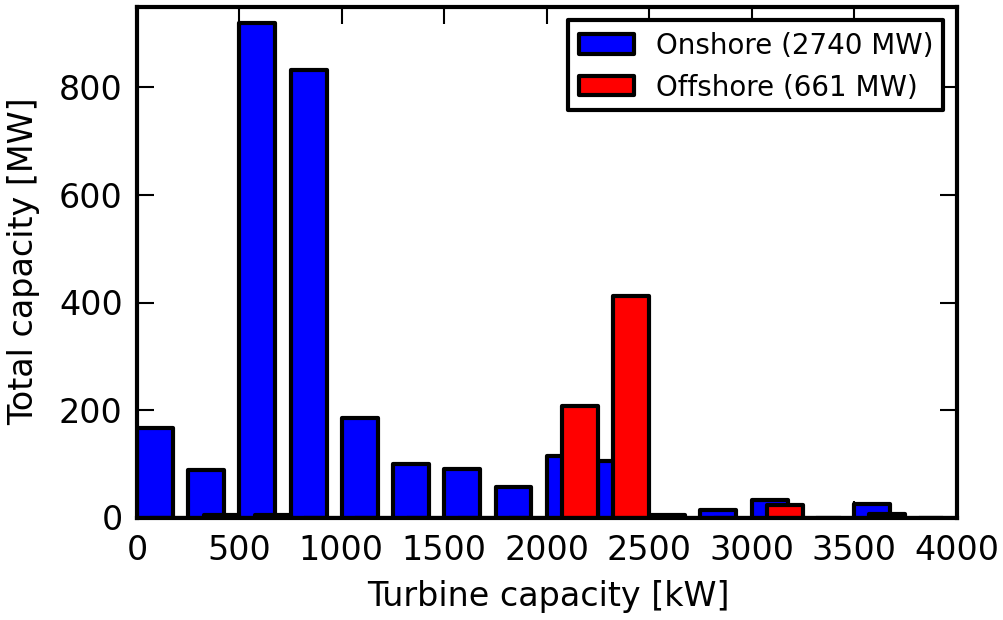}
		\caption{2010: Historical distribution of wind turbines.}
	\end{subfigure}
	
	\caption{Historical distribution of Danish wind turbines in the years (a) 2000, (b) 2005, and (c) 2010. To the left, the geographical distribution of total turbine rated capacity is shown. Each turbine is assigned to the nearest CFSR grid cell. To the right, the distribution of all individual turbine rated capacities is shown in bins of 250 kW.}
	\label{figure:WindLayout2000to2010} 
\end{figure*}

\begin{figure*}[tp!]
	\centering	

	\begin{subfigure}[b]{\linewidth}
		\includegraphics[width=0.49\linewidth]{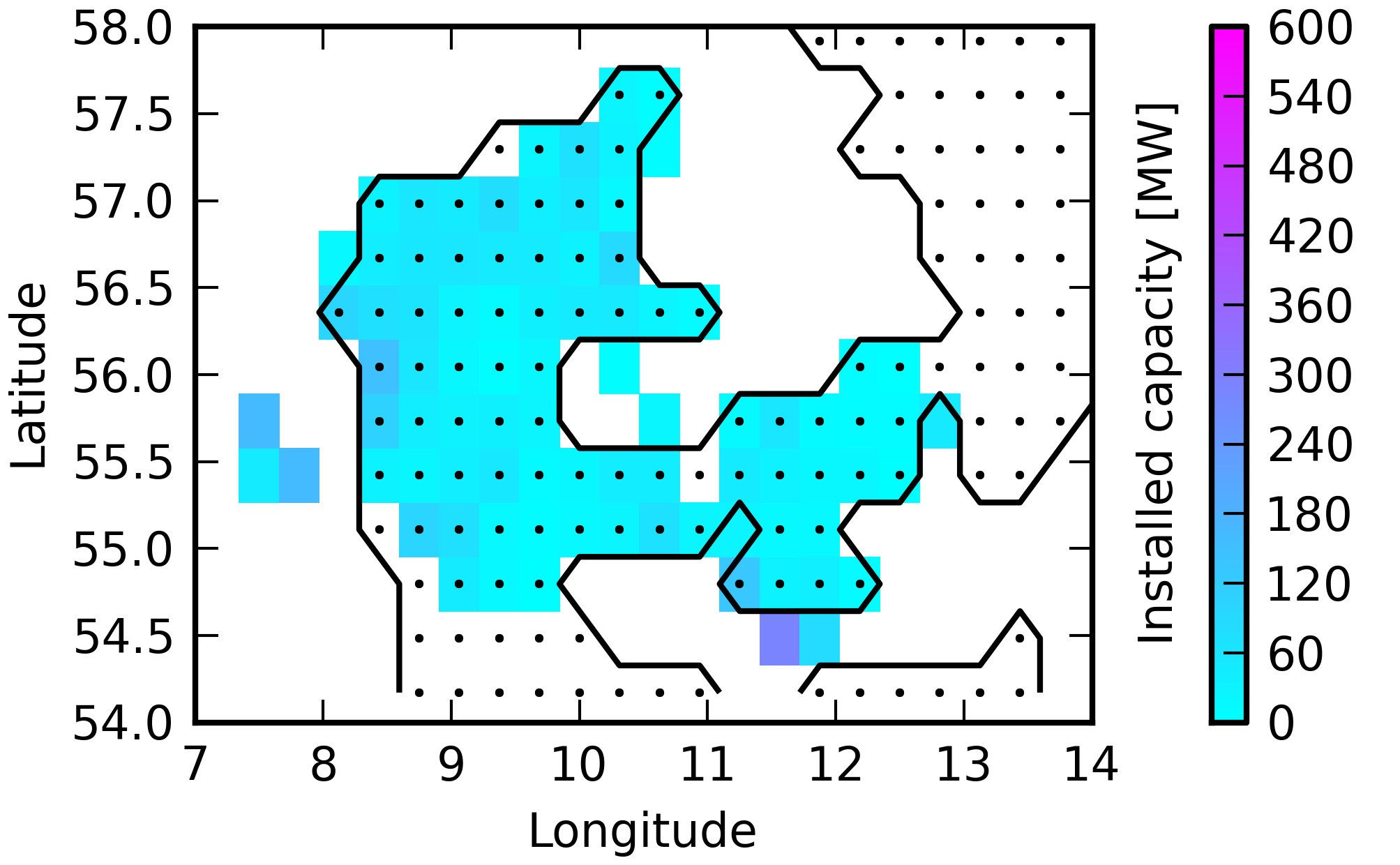}
		\includegraphics[width=0.49\linewidth]{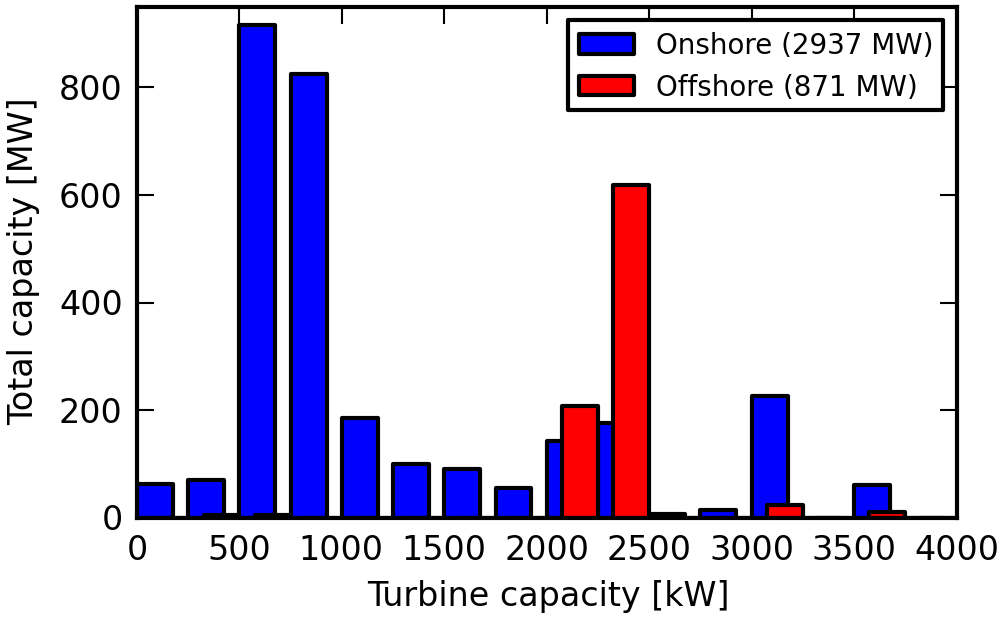}
		\caption{2012: Historical distribution of wind turbines.}
	\end{subfigure}		
		
	\begin{subfigure}[b]{\linewidth}
		\includegraphics[width=0.49\linewidth]{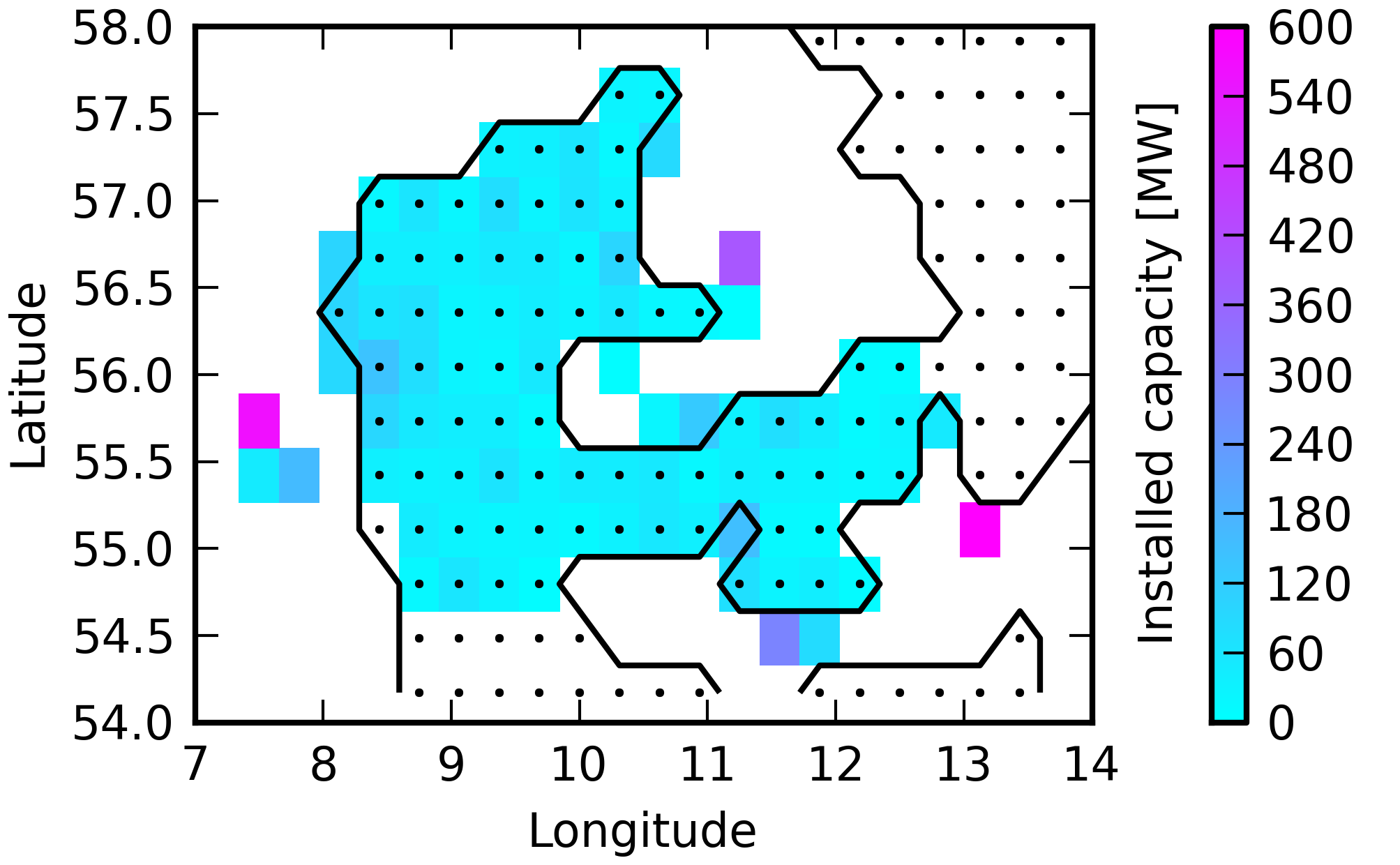}
		\includegraphics[width=0.49\linewidth]{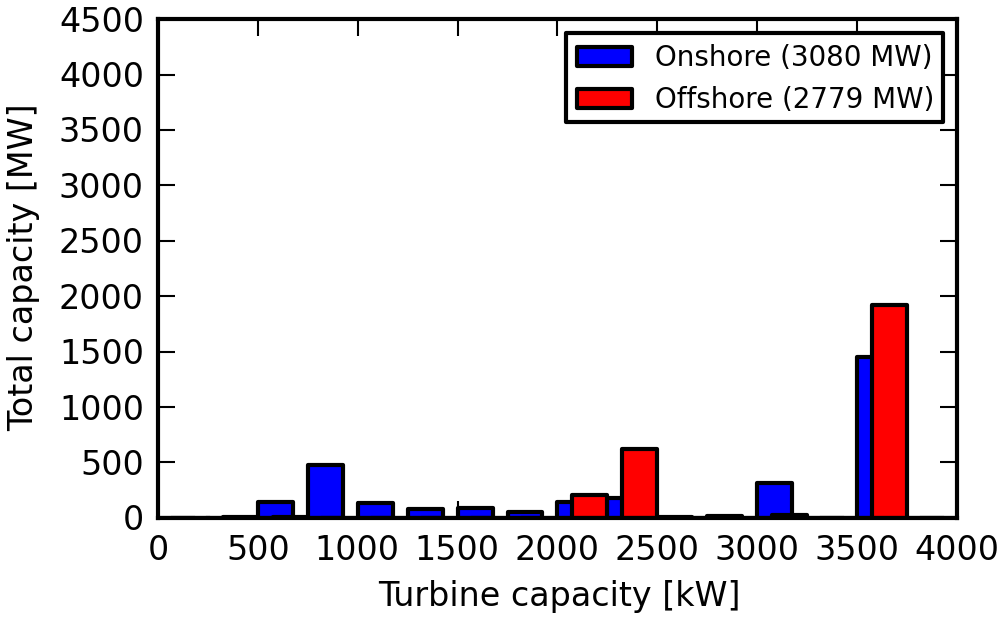}
		\caption{2020: Expected distribution of wind turbines.}
	\end{subfigure}
	
	\begin{subfigure}[b]{\linewidth}
		\includegraphics[width=0.49\linewidth]{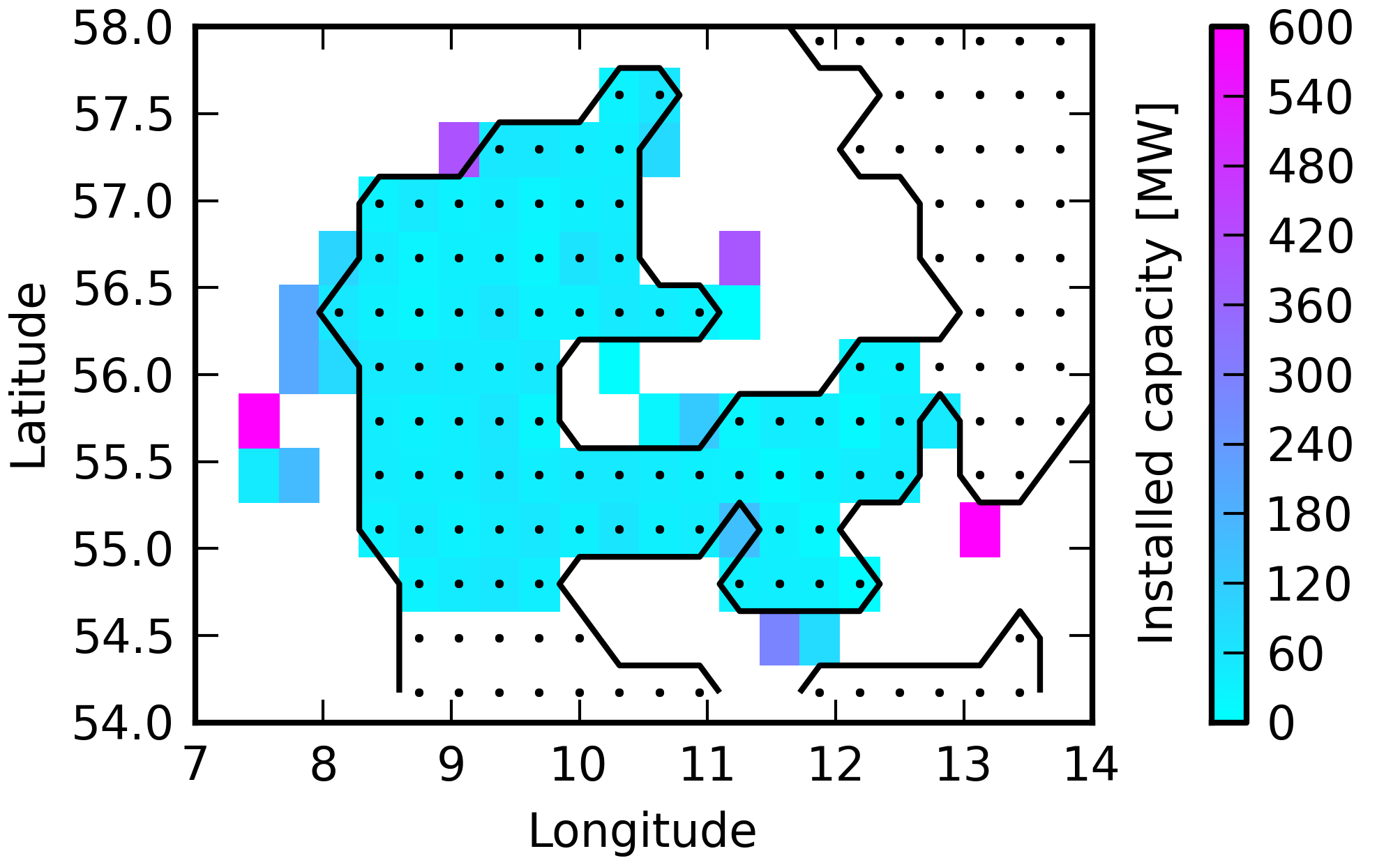}
		\includegraphics[width=0.49\linewidth]{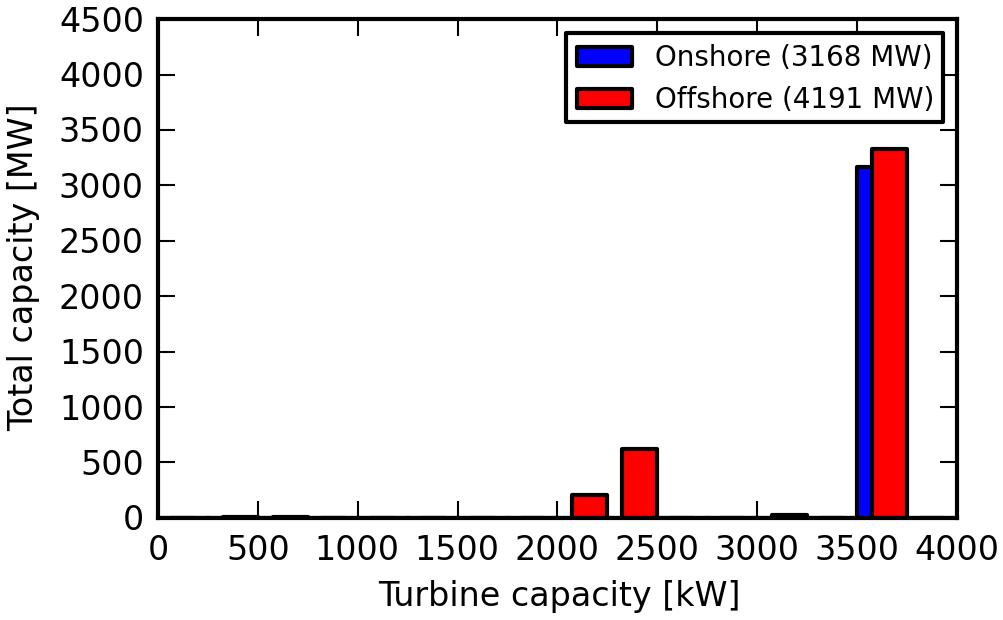}
		\caption{2035: Expected distribution of wind turbines.}
	\end{subfigure}
	
	\caption{Historical and expected distribution of Danish wind turbines in the years (a) 2012, (b) 2020, and (c) 2035. To the left, the geographical distribution of total turbine rated capacity is shown. Each turbine is assigned to the nearest CFSR grid cell. To the right, the distribution of all individual turbine rated capacities is shown in bins of 250 kW.}
	\label{figure:WindLayout2012to2035} 
\end{figure*}

In this paper, we discuss the historical years 2000, 2005, 2010, and 2012 and the future years 2020 and 2035 in detail. For these years, the geographical distribution of total turbine capacity as well as the distribution of the individual turbine capacities is shown in figures~\ref{figure:WindLayout2000to2010} and \ref{figure:WindLayout2012to2035}. The distributions are calculated for January 1$^\textup{st}$ of the model year.

In the year 2000, the turbine population consists almost exclusively of small onshore turbines with capacities less than 1,000~kW. This changes gradually towards 2012, where a number of turbines with capacities up to 3,600~kW are added onshore as well as offshore. Towards 2020, most of the small onshore turbines are expected to be replaced by large onshore turbines, and a significant amount of new offshore capacity is added. These trends are continued towards 2035, where all onshore turbines build before 2015 are assumed to have been replaced by new larger models.

\subsection{Historical wind power 2000 -- 2013}
Aggregated hourly wind power generation for Denmark has been downloaded from the Danish TSO Energinet.dk \cite{Energinet.dk:2013xy}. The data is updated continuously and dates back to the beginning of 2000. 

Here, the time series for the two Danish electricity system areas {DK1} and {DK2} have been combined and checked for obvious errors. This includes replacing missing or negative data with values from the following hour. The same is done for hours where the total production exceeds the installed capacity. Finally, the time stamps have been converted from local Danish time to {UTC} time, which does not include daylight saving hours. Only a few hours per year requires correction.

In the following sections, model time series are compared with historical data. For each model year in the period 1980 to 2035, the model is based on a static turbine distribution throughout all 32 weather years. The historical turbine distribution, on the other hand, is changing as old turbines are decommissioned and new turbines are grid connected. To partially correct for this effect, the historical wind power generation is normalised by the total installed rated capacity calculated with daily resolution. This is illustrated in Figure~\ref{figure:HistoricalWindCapacities}a where the hourly historical time series is a grey silhouette below the total installed capacity. In Figure~\ref{figure:HistoricalWindCapacities}b, the corresponding normalized time series is shown. Also shown is the average of the normalized time series for each calendar year. The average indicates the variation in average {CF} between years.

\subsection{{REatlas} wind conversion for Denmark}
Each model year $y$ between 1980 and 2035 is characterised by the historical or projected population of all Danish wind turbines on the first day of the year (cf. Section~\ref{section:turbineCapacityAndPlacement}). Onshore and offshore turbines are treated separately, and in both cases the turbines are divided into 250~kW wide categories by their capacity. The turbine power curve and hub height assigned to each category is summarised in Table~\ref{table:TurbineClasses}. In total, 12 onshore and 5 offshore categories in the interval 0 to 4,000~kW are used. Below, these categories are denoted by $X$.

\begin{table*}[bp!]
	\begin{tabular}{ l l l}
  		Capacity & Onshore & Offshore \\
		\hline
		0 -- 250 kW & Vestas V25 -- 200 kW (29~m) & - \\
  		250 -- 500 kW & NORDTANK -- 300 kW(31 m)  & - \\
 		750 -- 1000 kW & Vestas V47 -- 660 kW (45 m) & Vestas V39 -- 500 kW (41 m) \\
		1000 -- 1250 kW & Vestas V52 -- 850 kW (65 m) & - \\
		1250 -- 1500 kW & Nordex N60 -- 1300 kW (46 m)  & - \\
		1500 -- 1750 kW & Vestas V66 -- 1.65 MW (67 m) & - \\
		1750 -- 2000 kW & Vestas V66 -- 1.75 MW (67 m) & - \\
		2000 -- 2250 kW & Vestas V80 -- 2 MW (80 m)  & Vestas V80 - 2 MW (70 m)  \\
		2250 -- 2500 kW & Siemens SWT 2.3 -- 93 (80 m)  & Siemens SWT 2.3 -- 93 (68 m) \\
		2500 -- 2750 kW & - & - \\
		2750 -- 3000 kW & NEG Micon -- 2750 kW (60 m) & - \\
		3000 -- 3250 kW & Vestas V90 -- 3 MW (80 m)  & Vestas V90 -- 3 MW (70 m) \\
		3250 -- 3500 kW & - & - \\
		3500 -- 3750 kW & Siemens SWT 107 3.6 MW (90 m) & Siemens SWT 107 3.6 MW (82 m)  \\
		3750 -- 4000 kW & - & - \\
	\end{tabular}
	\caption{Wind turbine classes. The hub height is given in brackets after the turbine name. }
	\label{table:TurbineClasses}
\end{table*}

A capacity layout map $\{C_{X,n} \}_y$ similar to those shown in Figure~\ref{figure:WindLayout2000to2010} and \ref{figure:WindLayout2012to2035} is then created for each turbine category by assigning individual onshore (offshore) turbines to the nearest onshore (offshore) {CFSR} grid point $n$. For each capacity map, the corresponding power curve and hub height are then used to model the normalised hourly wind power generation $W_{X,n}$ for the 32 weather years in the {REatlas} (cf. Section~\ref{section:REatlas}). Finally, the time series of all categories are aggregated and added together to produce a model wind power time series $W_y$ for the selected model year.

\begin{equation}
	W_y(t) = \sum_{X,n} C_{X,n}\, W_{X,n}(t)\,,
\end{equation}
where $C_{X,n}\in\{C_{X,n} \}_y$.

\subsection{Calibration of the {REatlas} wind conversion for Denmark}
\begin{figure}[tp!]
	\centering	

	\begin{subfigure}[b]{\columnwidth}
		\includegraphics[width=\linewidth]{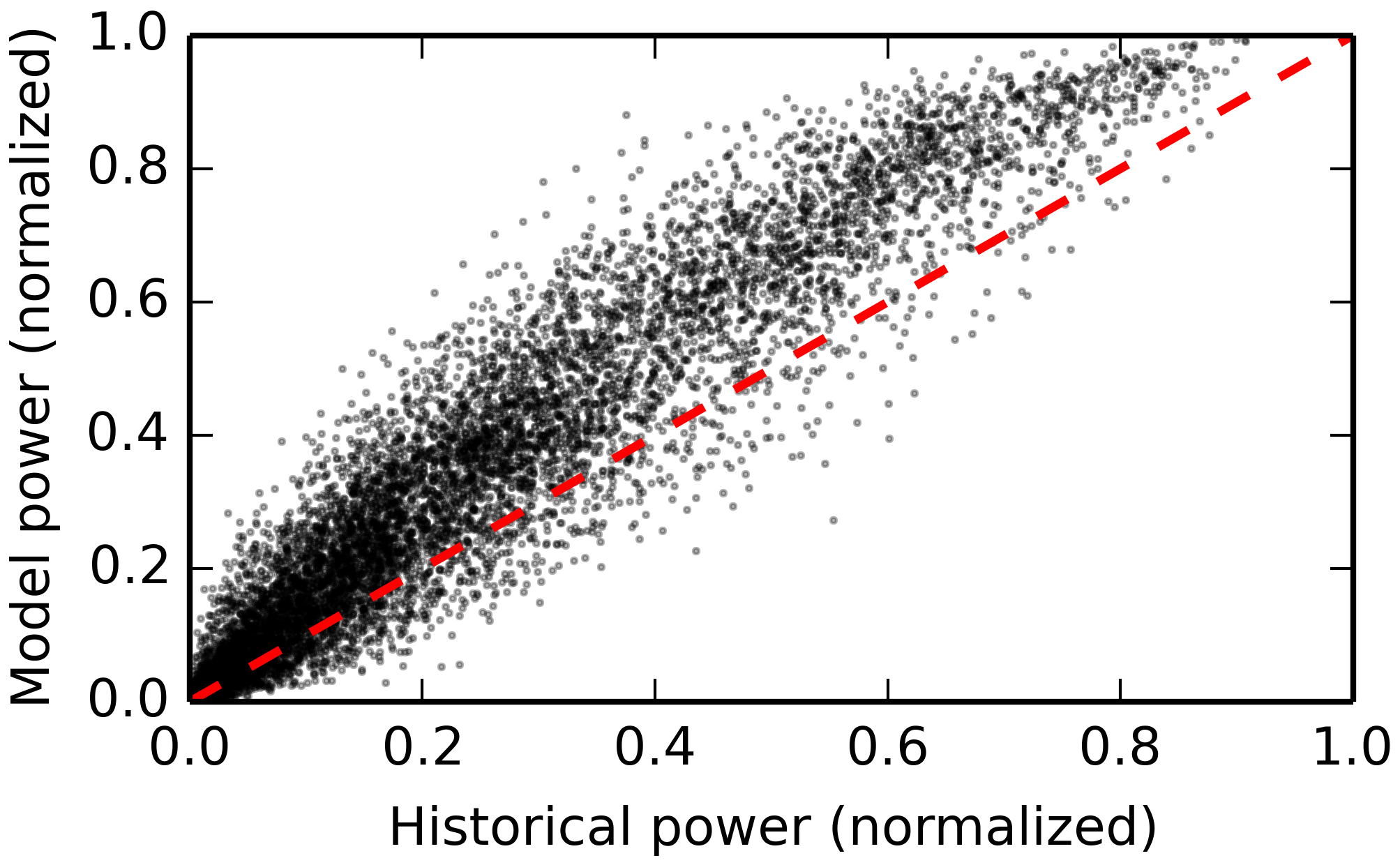}
		\caption{2010: Initial model vs. historical wind power.}
	\end{subfigure}	
	\begin{subfigure}[b]{\columnwidth}
		\includegraphics[width=\linewidth]{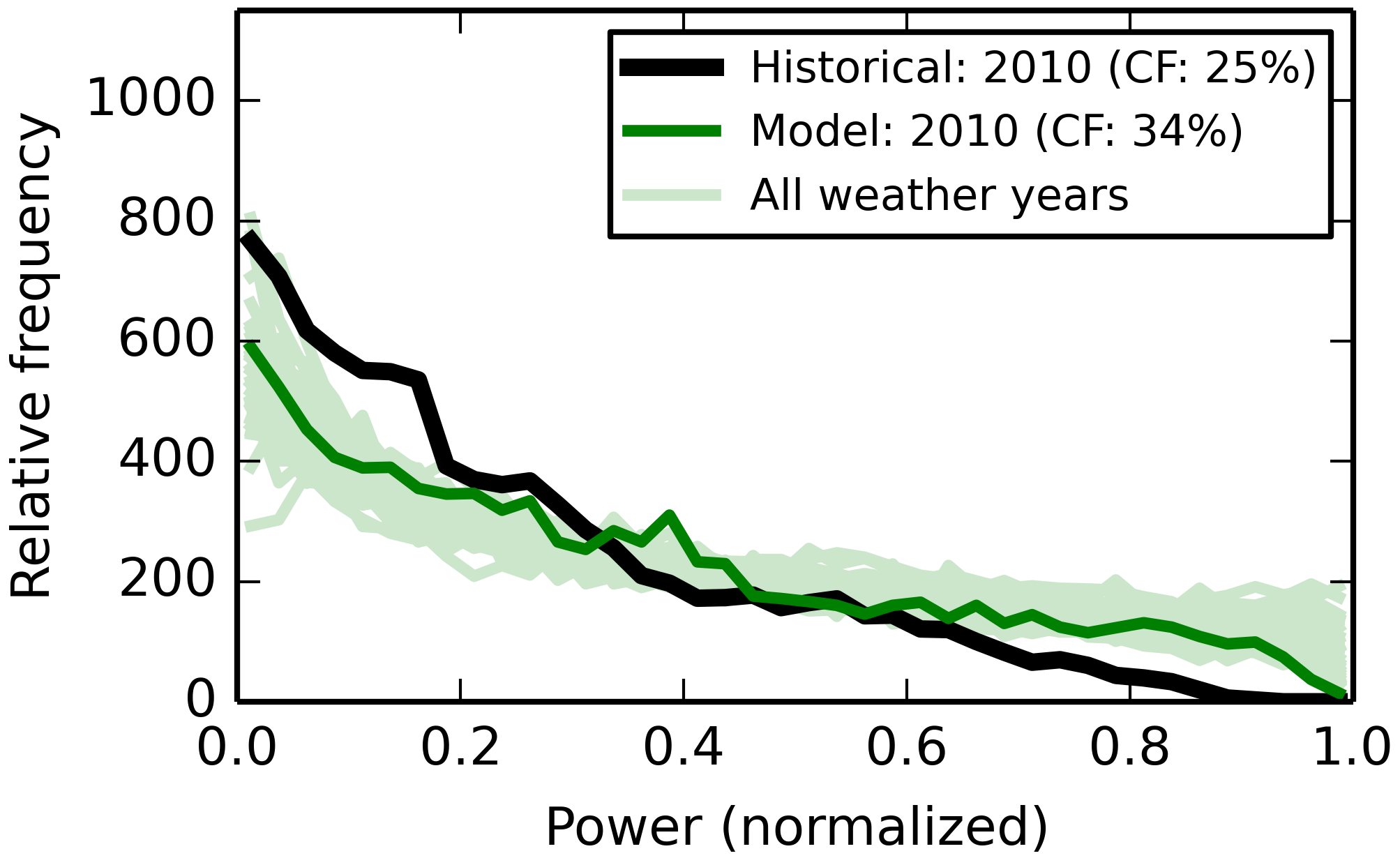}
		\caption{2010: Annual distribution of wind power.}
	\end{subfigure}	
	
	\caption{Comparison between initial model and historical wind power generation for the model year 2010. In the initial model, the turbine power curves specified by the manufacturer are used directly in the wind conversion. (a) Modelled hourly wind power vs. historical wind power generation. All values are normalised to the total installed capacity, and the dashed red line indicates a 1:1 match. (b) Annual distribution of the hourly wind power generation. The distributions are: Historical data for 2010 (black) and initial model output for the weather year 2010 (green). Model results for all 32 weather years are shown as overlapping light green traces.}
	\label{figure:InitialMatchWind} 
\end{figure}

Initially, the turbine power curves as specified from the manufacturer are used directly in the wind conversion algorithm (see Section~\ref{section:windConversion}). The outcome of this model calculation is compared to historical data for the model year 2010 in Figure~\ref{figure:InitialMatchWind}. In Figure~\ref{figure:InitialMatchWind}a, the model output from the weather year 2010 and historical wind generation from the same year is compared directly hour-by-hour. The figure shows a strong correlation between the two, but the model is clearly biased towards higher production. In addition, the figure shows significant variation in the model output for a given historical value. Most pronounced in the central part of the figure. The latter effect is to be expected, however, since small differences between model and real wind speeds can be amplified by the $\nu^3$ behaviour of the middle part of the power curve. 

In Figure~\ref{figure:InitialMatchWind}b the annual distribution of wind power generation for model and historical data are compared. These distributions do not contain information about the temporal correlation between the hours, and at least part of the variation in model output for a given historical value is expected to average out. The two distributions are qualitatively similar, but the model bias towards high output results in a clear increase in average capacity factor (CF). For the weather year 2010, the model predicts a CF of 34\% whereas the historical CF was 25\%.\\

In \cite{Norgaard:2004uq}, a smoothing of the single turbine power curve used above is proposed in order to account for the effects of variations in the wind speed within each grid cell and each time step. In addition, all turbines may not perform exactly as specified by the manufacturer and for a large population such as for Denmark, some turbines will always be unavailable due to planned or unplanned maintenance. To capture these effects, we introduce a simple heuristic smoothing function which is applied to the single turbine power curves before they are used in the wind conversion. The smoothing function is based on, but not identical to the one of \cite{Norgaard:2004uq}. 

The modified power curve $P_\textup{mod}$ is calculated for all wind speeds $\nu_0$ below the cut-out speed as the convolution:
\begin{equation}\label{equation:smoothingFunction}
	P_\textup{mod}(\nu_0) = \eta \int_0^\infty  P_{0}(\nu) \Ker(\nu_0,\nu - \Delta \nu) \textup{d}\nu\,.
\end{equation}
Here $\eta$ is the average wind turbine fleet availability, $\Delta\nu$ an offset, and  $\Ker$ a smoothing function defined as follows:
\begin{equation}
	\Ker(\nu_0,\nu - \Delta v)  = \frac{1}{\sqrt{2\pi\sigma_{0}^2}}\, e^{-\frac{\left(\nu_0 + \Delta \nu - \nu\right)^2}{2\sigma_{0}^2}}\,.
\end{equation}
The integral of the kernel is normalized to unity and its functional form is a Gaussian with standard deviation $\sigma_0$ and mean value $\nu_0+\Delta\nu$. 

\begin{figure}[tp!]
	\centering	
	
	\includegraphics[width=0.99\linewidth]{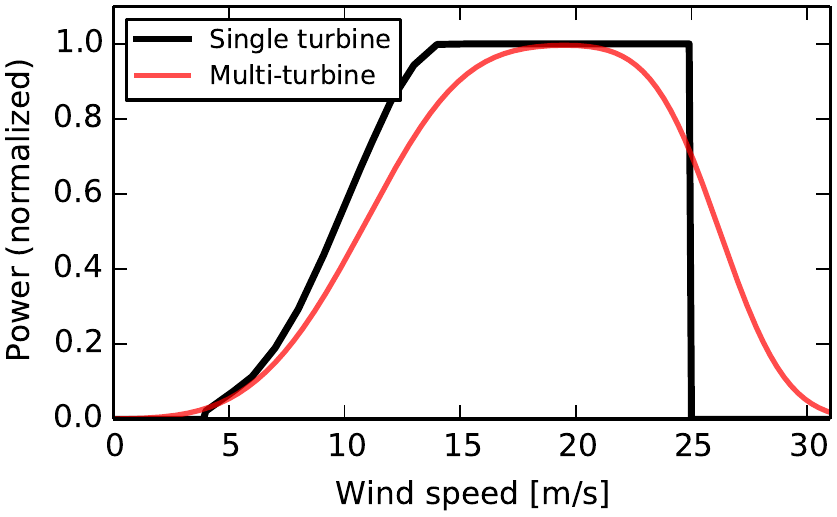}
	
	\caption{Example of a power curve with and without the modification Eq.~\eqref{equation:smoothingFunction} applied. The original power curve is taken from the data sheet of a Vestas V90 3~MW turbine (black line). The modified power curve (red line) is calculated with the standard deviation $\sigma_0=2.29$~m/s and the offset $\Delta\nu=1.27$~m/s.}
	\label{figure:PowerCurveModification} 
\end{figure}

The three parameters of the smoothing function are assumed to be identical for all turbine categories (see Table~\ref{table:TurbineClasses}) and they are determined by comparing data for the model and weather year 2010 with historical wind power for the same year. This is done by non-linear numerical least square minimization of the difference between model and historical data. The optimal values are found to be about $1.27$~m/s for the offset $\Delta\nu$, and about 2.29~m/s for the standard deviation $\sigma_0$. The fleet availability $\eta$ is found to be 95\%. A number, which roughly corresponds to numbers in the literature \cite{Energistyrelsen:2012uq}. An example of a modified power curve with these parameters is shown in Figure~\ref{figure:PowerCurveModification}. Note that the numerical parameter optimization is greatly facilitated by the speed at which repeated conversions are performed with the {REatlas}.

\begin{figure*}[tp!]
	\centering	
	
	\begin{subfigure}[b]{0.49\linewidth}
		\includegraphics[width=\linewidth]{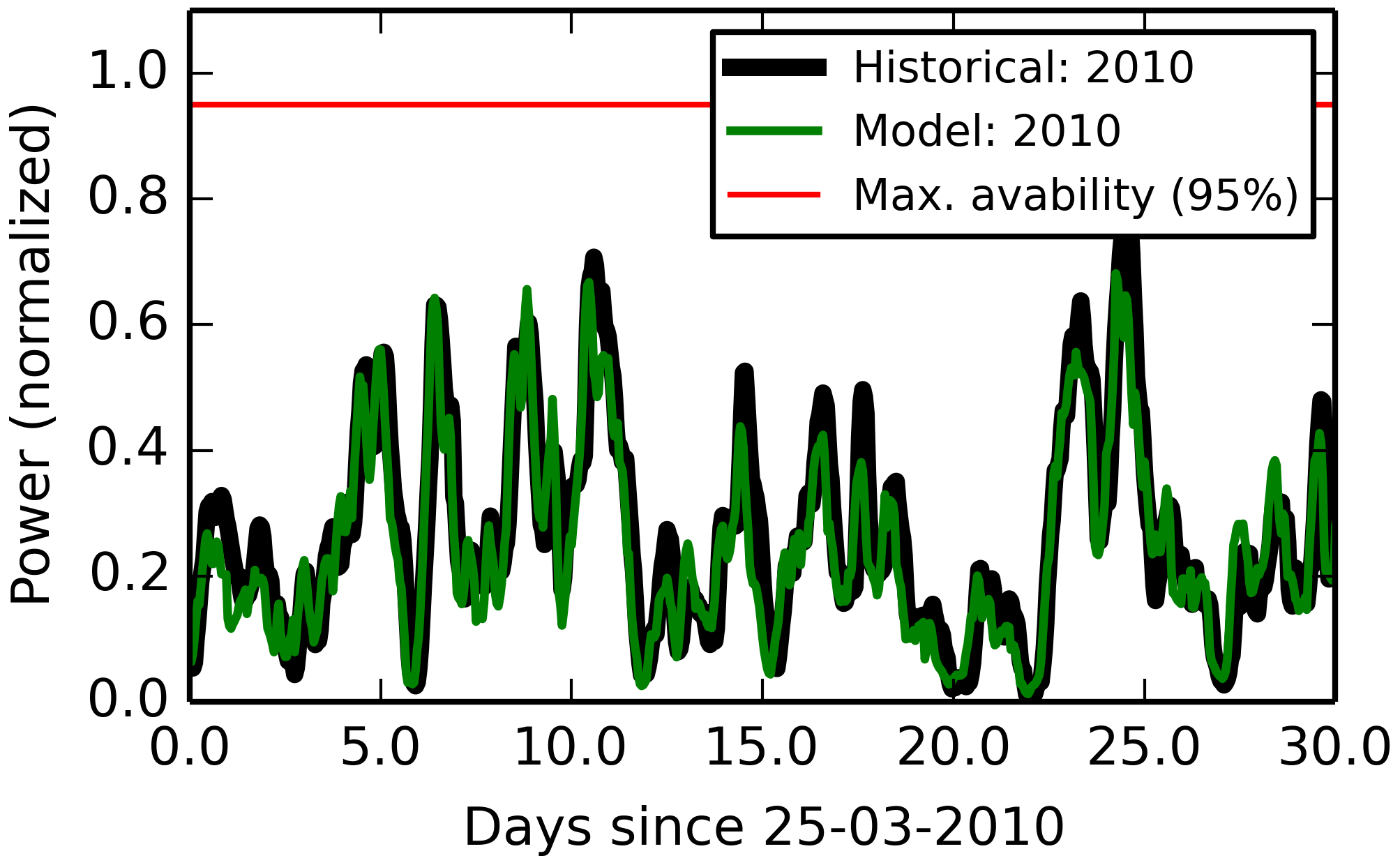}
		\caption{Sample time series.}
	\end{subfigure}
	\begin{subfigure}[b]{0.49\linewidth}
		\includegraphics[width=\linewidth]{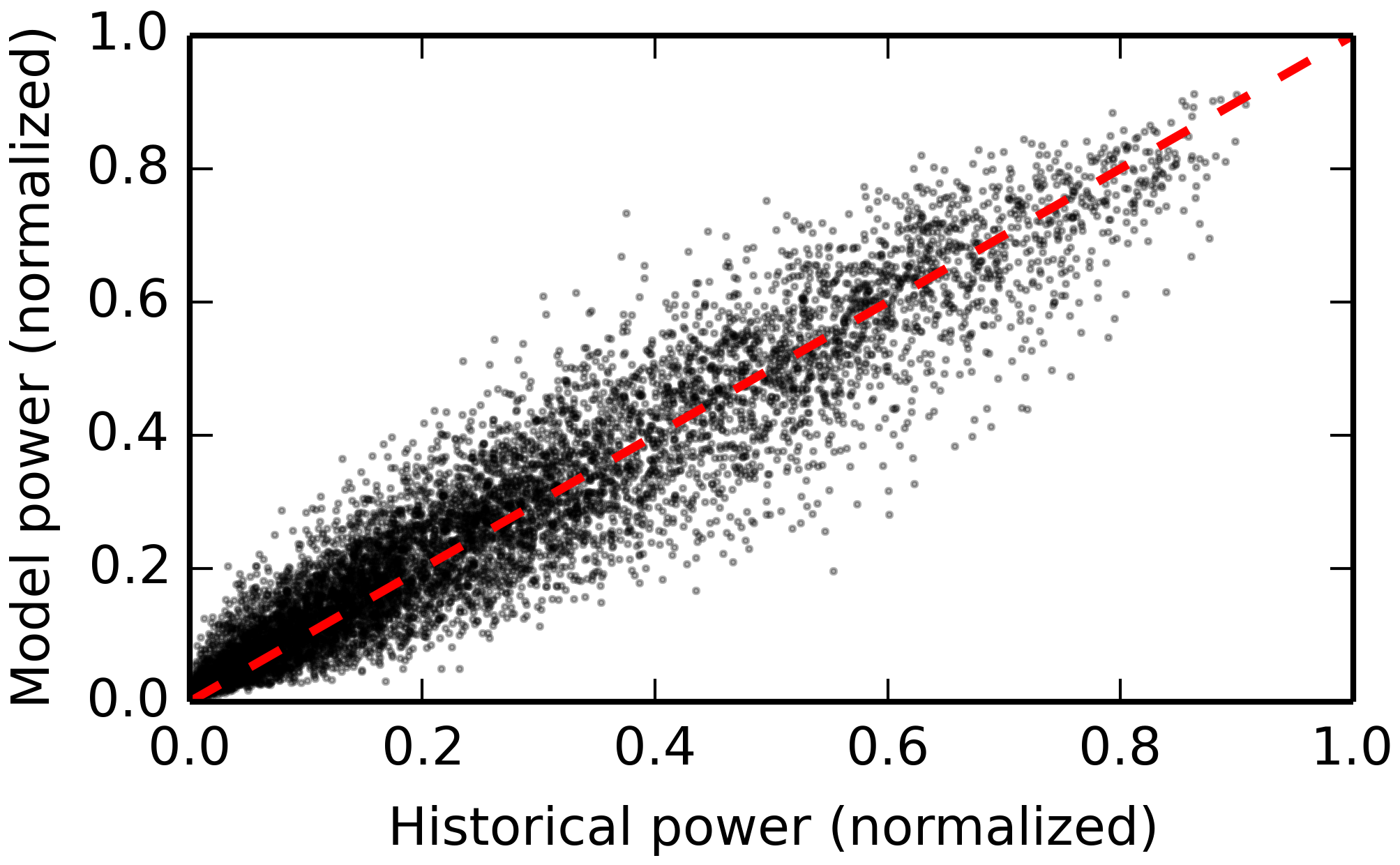}
		\caption{Model vs. historical data.}
	\end{subfigure}
	\begin{subfigure}[b]{0.49\linewidth}
		\includegraphics[width=\linewidth]{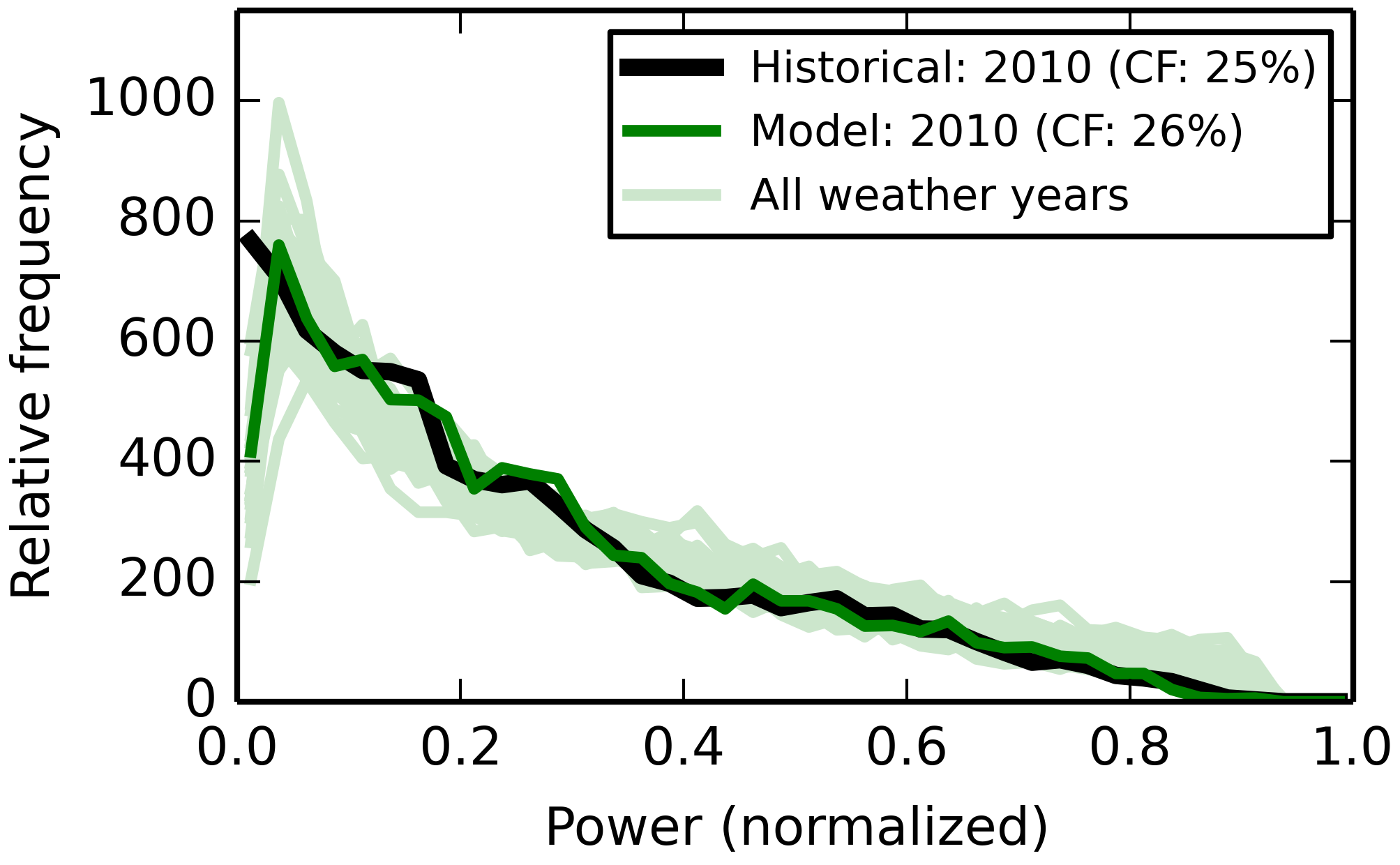}
		\caption{Distribution of power generation.}		
	\end{subfigure}
	\begin{subfigure}[b]{0.49\linewidth}
		\includegraphics[width=\linewidth]{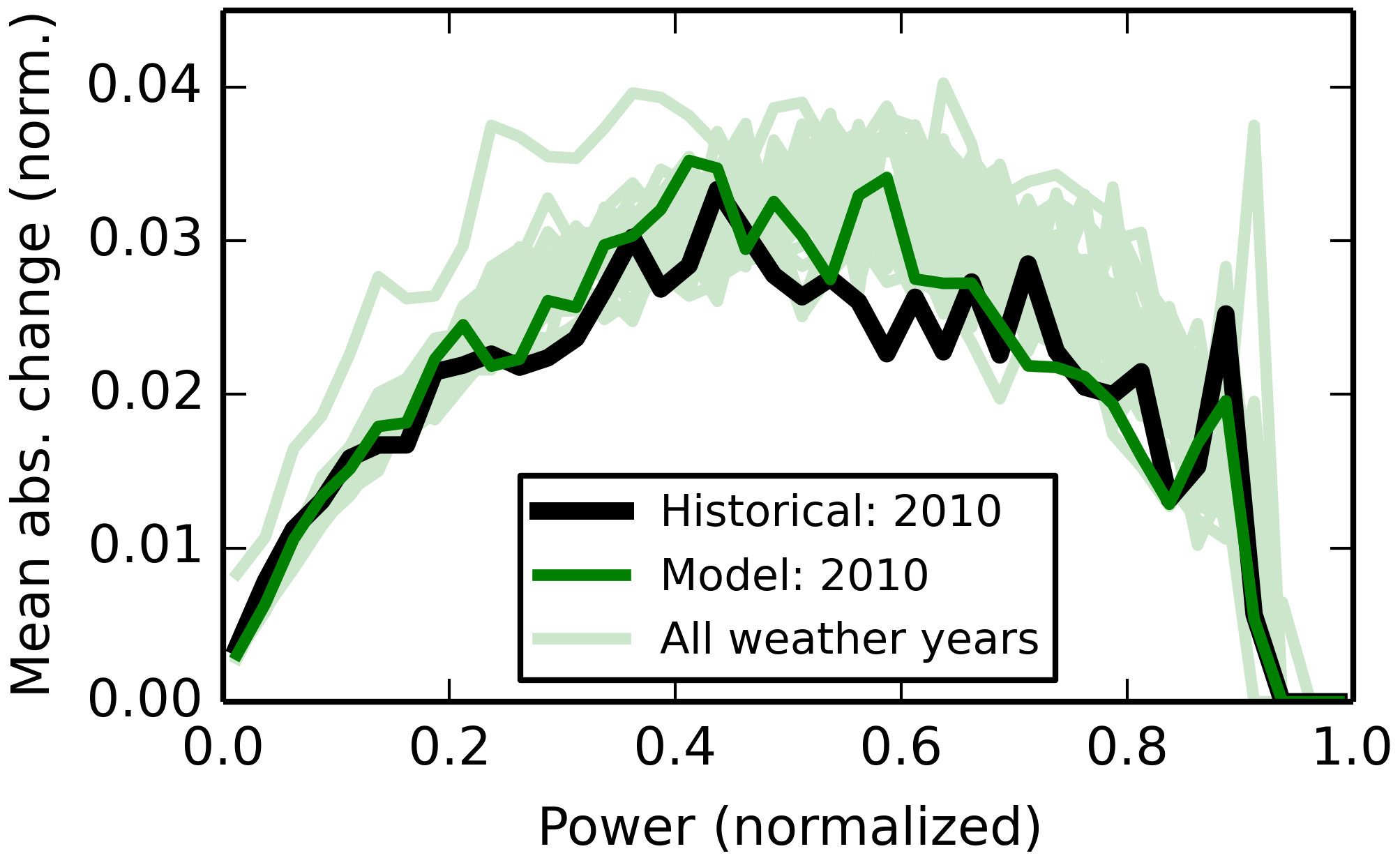}
		\caption{Mean abs. change vs. power generation.}
	\end{subfigure}
	\begin{subfigure}[b]{0.49\linewidth}
		\includegraphics[width=\linewidth]{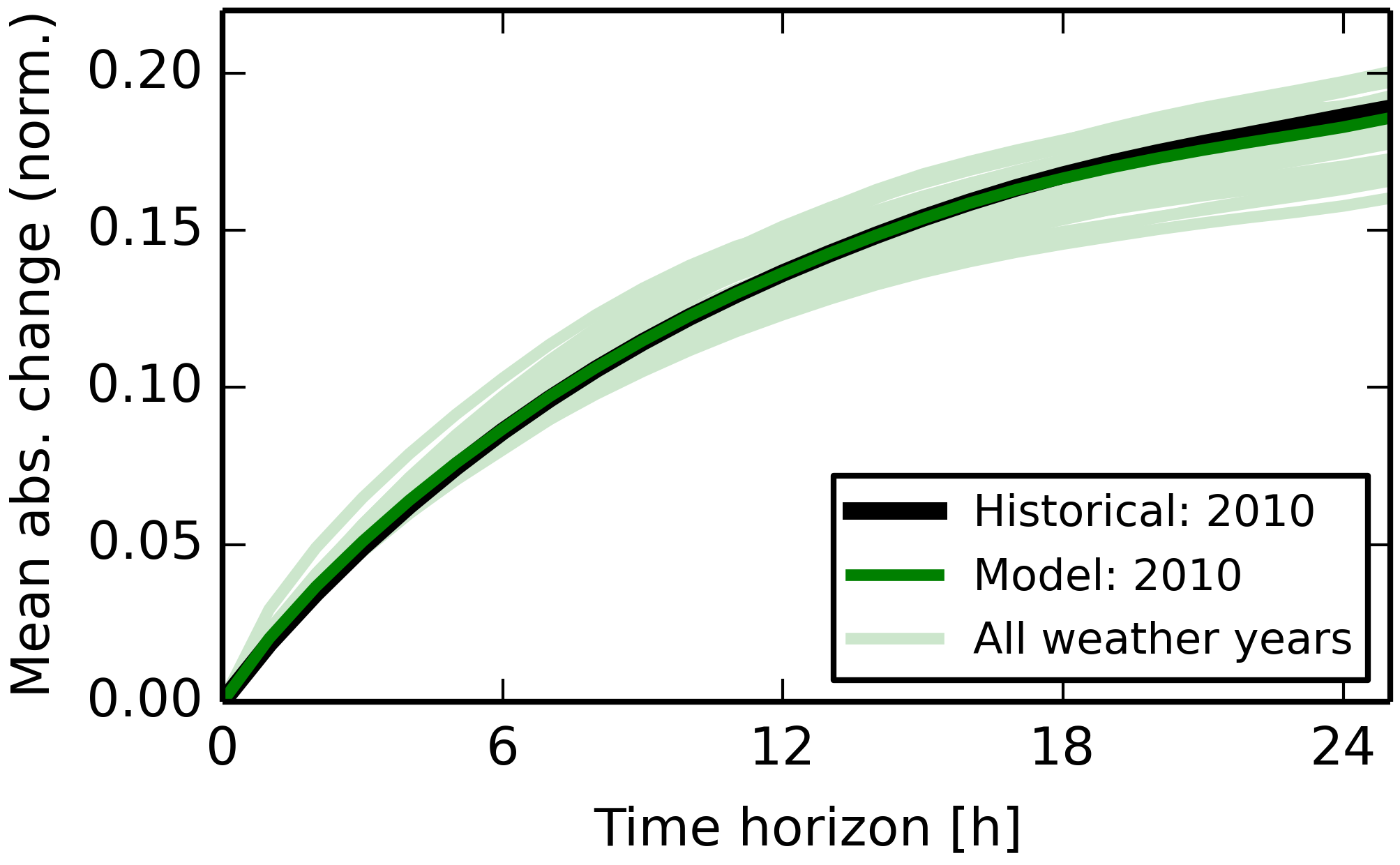}
		\caption{Mean abs. change vs. time horizon.}
	\end{subfigure}
	\begin{subfigure}[b]{0.49\linewidth}
		\includegraphics[width=\linewidth]{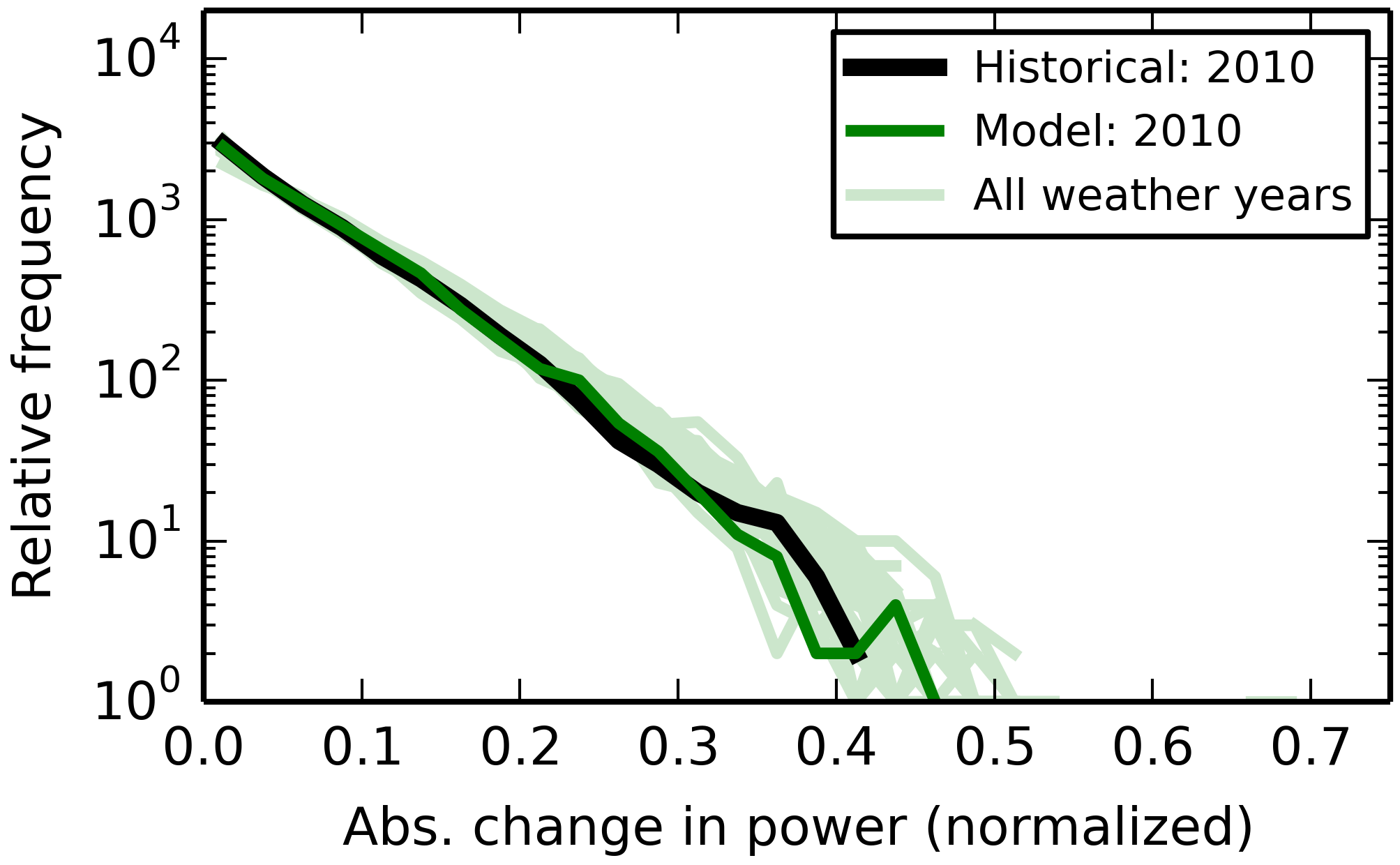}
		\caption{Distribution of mean abs. change (4~h).}
	\end{subfigure}	
	\caption{Comparison between hourly model wind power (green) and historical data (black) for 2010 using the power curve modification Eq.~\eqref{equation:smoothingFunction} with standard deviation $\sigma_0=2.29$~m/s, offset $\Delta\nu=1.27$~m/s, and availability $\eta=95\%$. The high-lighted model data is based on the weather year 2010. Model output for all 32 weather years is shown as overlapping light green traces.}
	\label{figure:Validation2010} 
\end{figure*}

Figure~\ref{figure:Validation2010} shows a comprehensive comparison between model data calculated with the modified power curve and historical data for the year 2010. In Figure~\ref{figure:Validation2010}a and \ref{figure:Validation2010}b, model and historical data are compared directly. Figure~\ref{figure:Validation2010}b is similar to Figure~\ref{figure:InitialMatchWind}a, and it is evident that the individual hours are now distributed more symmetrically around the identity map. 

The figures~\ref{figure:Validation2010}c to \ref{figure:Validation2010}f compare different characteristics of the model and the historical wind power time series. The figures are inspired by similar figures found in \cite{Sturt:2011bh}, where it is argued that these characteristics are of particular importance to the integration of wind in the electricity system. Here, a good match between model and historical data for the year 2010 is found in all four plots. As shown in Figure~\ref{figure:Validation2010}c, the model distribution of hourly wind power generation now closely match the historical distribution except for very low outputs, where the model tends to underpredict the number of hours. The average CF for the model data is 26\%, which is close to the historical value of 25\%. 

Figures~\ref{figure:Validation2010}d to \ref{figure:Validation2010}f are concerned with the change in wind power generation between hours. Figure~\ref{figure:Validation2010}d shows the mean absolute change between consecutive hours as a function of power generation. Figure~\ref{figure:Validation2010}e shows the mean absolute change between hours separated by between 1 and 24 hours, and Figure~\ref{figure:Validation2010}f shows the relative frequency of absolute change in power generation for hours separated by 4 hours. Note that the latter exhibits exponential behaviour, i.e. the probability of a certain absolute change decreases exponentially with its magnitude.

\subsection{Results for selected years}\label{section:SelectedYearsWindPower}
\begin{figure*}[p!]
	\centering	

	\begin{subfigure}[b]{0.49\linewidth}	
	\includegraphics[width=\linewidth]{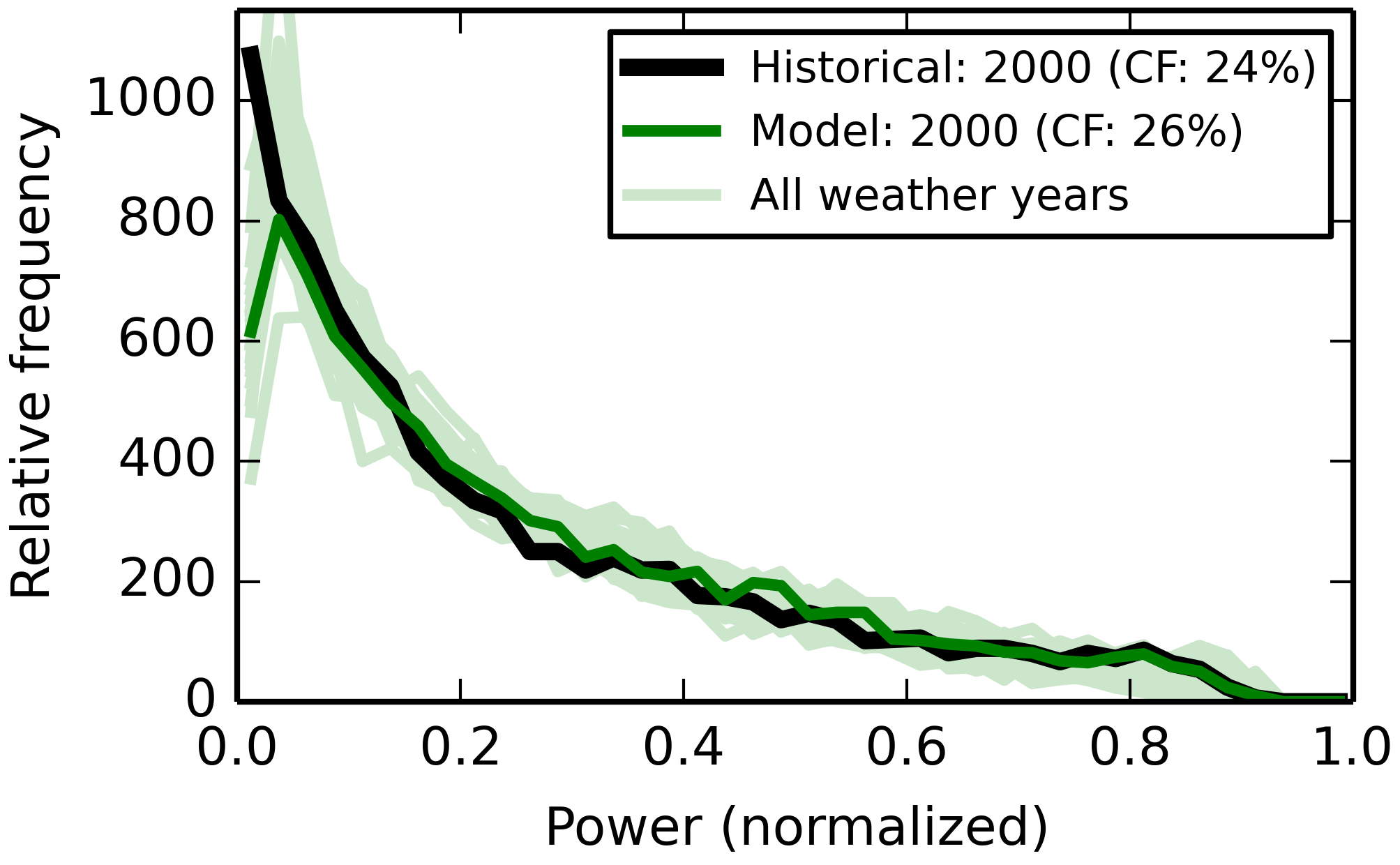}
		\caption{Model year: 2000}
	\end{subfigure}	
	\begin{subfigure}[b]{0.49\linewidth}
	\includegraphics[width=\linewidth]{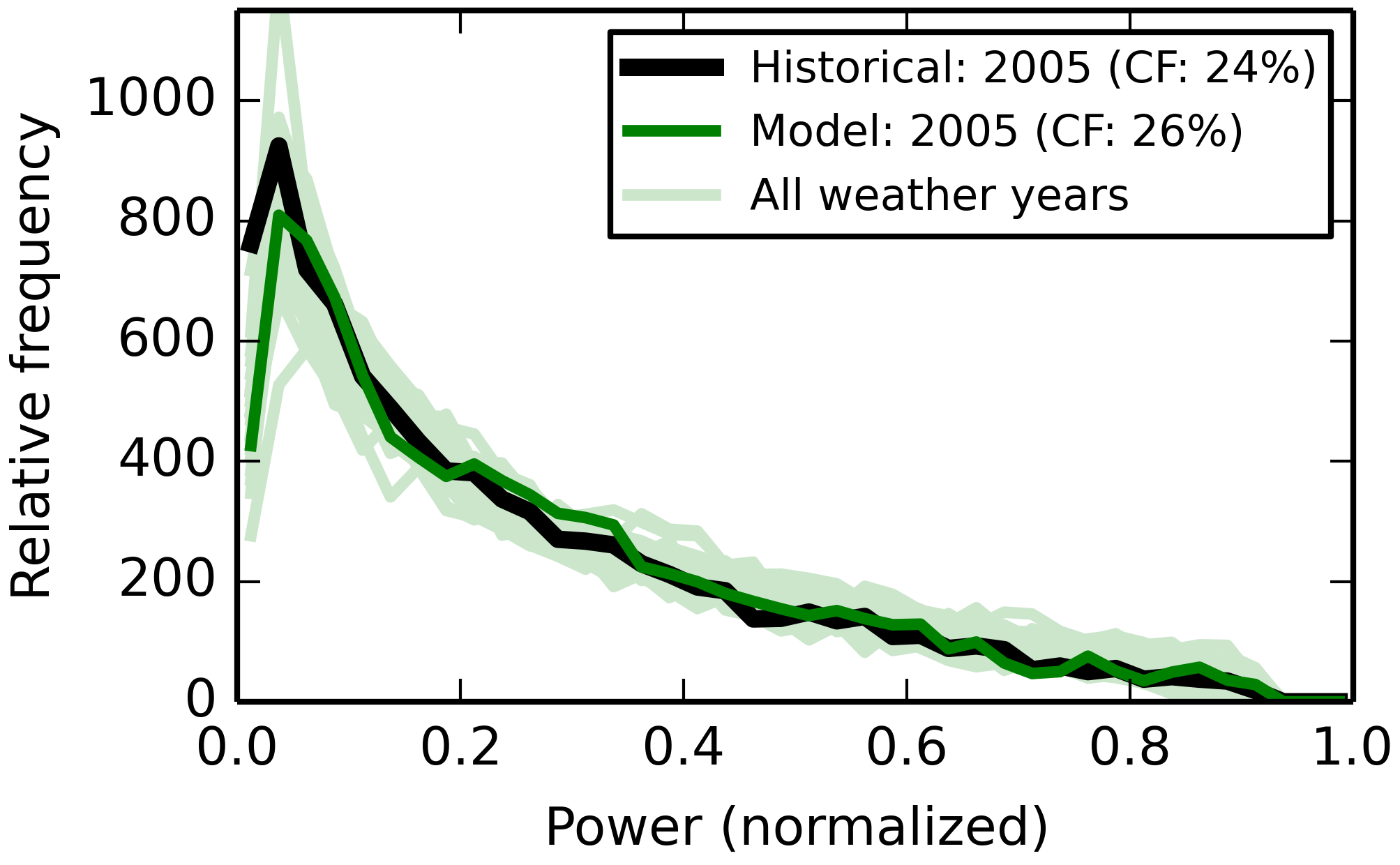}
		\caption{Model year: 2005}
	\end{subfigure}	
	\begin{subfigure}[b]{0.49\linewidth}
	\includegraphics[width=\linewidth]{{plot_scenario_Denmark_2010-01-01_smooth_2285_offset_1268_availability_95_freq_vs_power}.png}
		\caption{Model year: 2010}
	\end{subfigure}	
	\begin{subfigure}[b]{0.49\linewidth}
	\includegraphics[width=\linewidth]{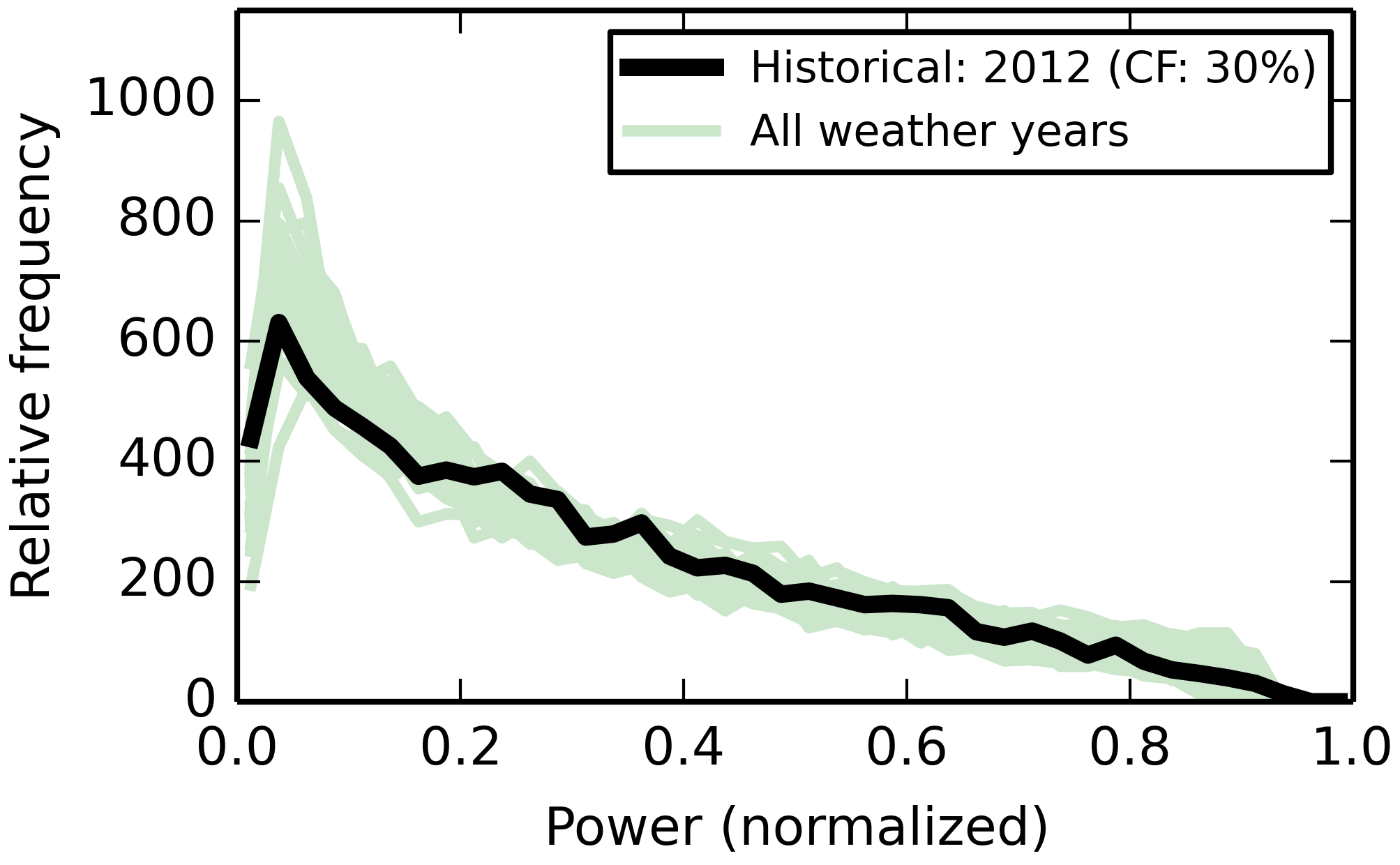}
		\caption{Model year: 2012}
	\end{subfigure}	
	\begin{subfigure}[b]{0.49\linewidth}
		\includegraphics[width=\linewidth]{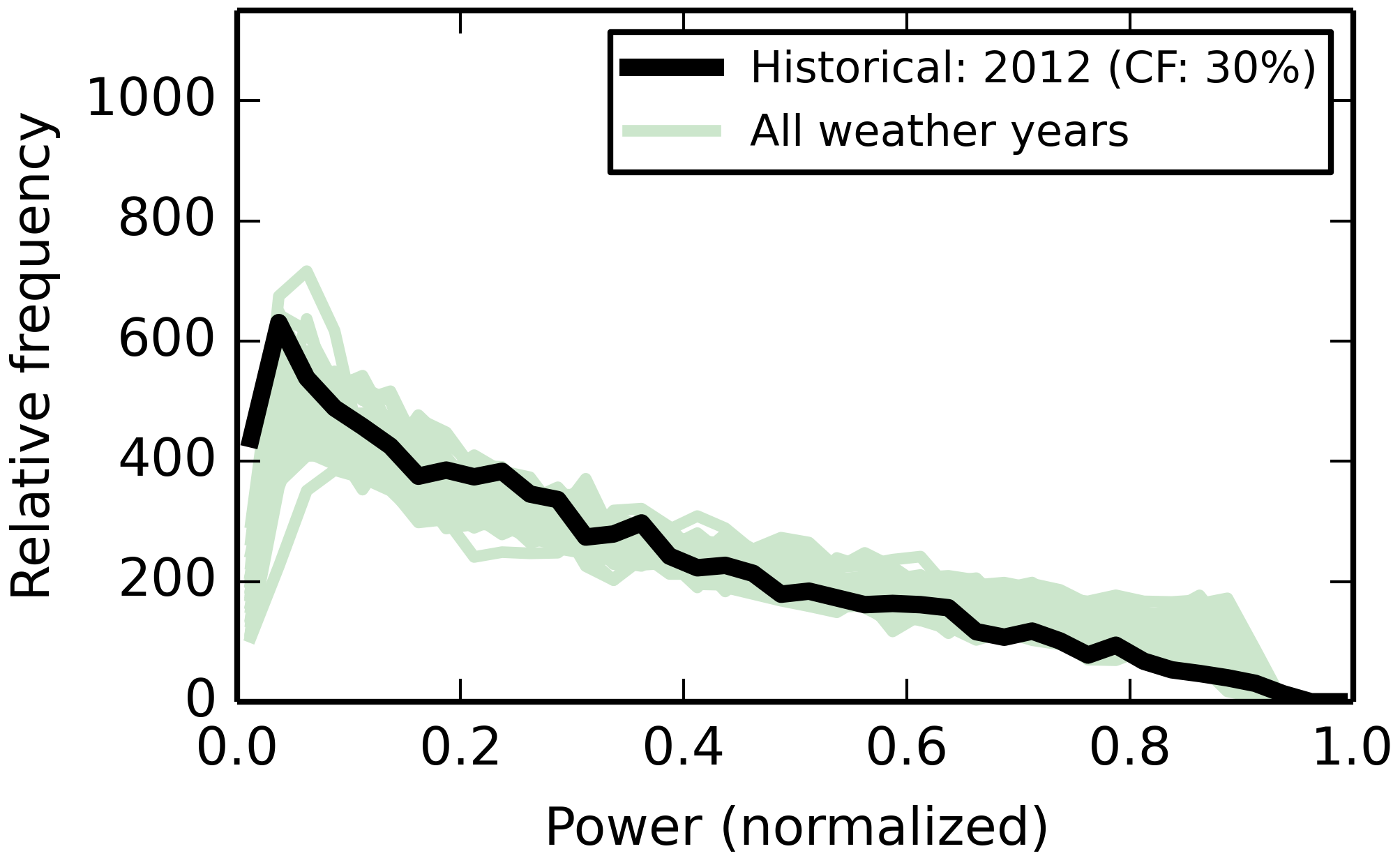}
		\caption{Model year: 2020}
	\end{subfigure}	
	\begin{subfigure}[b]{0.49\linewidth}
		\includegraphics[width=\linewidth]{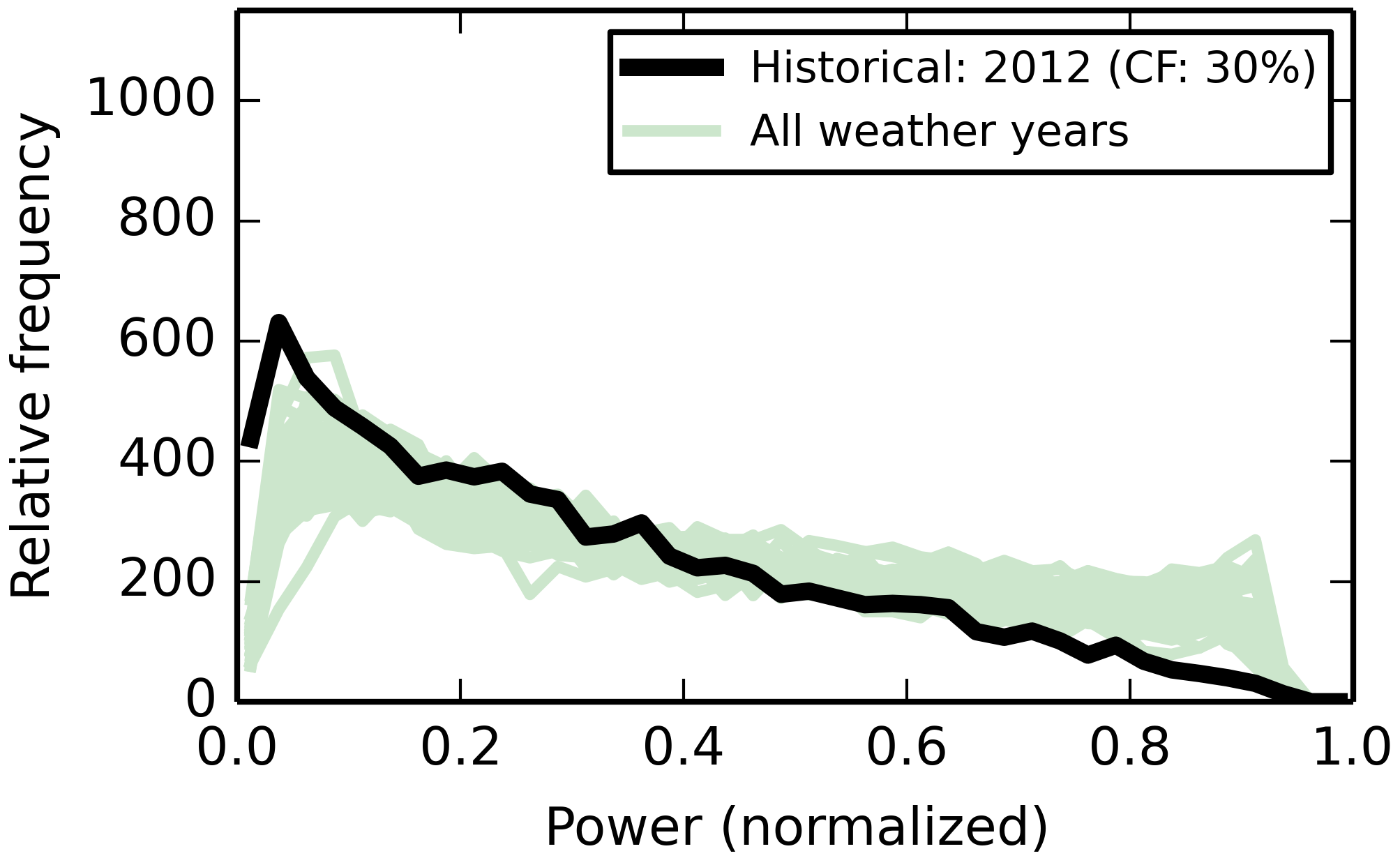}
		\caption{Model year: 2035}
	\end{subfigure}	
	\caption{Comparison between the annual distribution of hourly model wind power and historical data. For the historical years 2000 to 2010, historical data for the year (black) is compared to model data from the matching weather year. Model results for all 32 weather years are shown as overlapping light green traces. For the years 2012, 2020, and 2035, matching weather years are not available, and historical data is always from 2012. In the model calculations, the model year determines the population of wind turbines, and the power curve modificationEq.~\eqref{equation:smoothingFunction} is used with standard deviation $\sigma_0=2.29$~m/s, offset $\Delta\nu=1.27$~m/s, and availability $\eta=95\%$.}
	\label{figure:Validation2000_2035} 
\end{figure*}

The optimal parameters for the modified power curve Eq.~\eqref{equation:smoothingFunction} are determined for 2010 as described in the previous section. In this section, these parameters are used to calculate model wind power time series for the historical years 2000, 2005, 2010 and 2012 as well as for the future years 2020 and 2035. The results of these calculations are summarised in Figure~\ref{figure:Validation2000_2035} where the distribution of hourly power generation is shown.

For the historical years 2000, 2005 and 2010, we are able to make a direct comparison of historical power generation with model data based on the same model and weather year. As shown in the figures~\ref{figure:Validation2000_2035}a to \ref{figure:Validation2000_2035}c, the model distributions agree well with the historical distributions except for very low power outputs. The average CF of the model is within a few percentage points of the historical value.

For the historical year 2012, weather data is not currently available in the {REatlas} (cf. Section~\ref{section:weatherData}), which means that a direct comparison is not possible. Instead, the individual weather years are compared to the historical power generation of 2012. As shown in Figure~\ref{figure:Validation2000_2035}d, a reasonable agreement is achieved. Compared with the previous historical years, the average CF is higher for both model and historical data. This is consistent with the shift towards larger and more offshore turbines.

For the future years 2020 and 2035, historical data from 2012 is plotted for comparison. As shown in Figures~\ref{figure:Validation2000_2035}e and \ref{figure:Validation2000_2035}f, the number of hours with generation above 50\% of the installed capacity increase for both of these years as old onshore turbines are replaced with larger units and as more offshore turbines are added. This is reflected in an increase in the average CF, which is discussed further in the following section.

\subsection{Development of wind power characteristics in the future}\label{section:FutureCharacteristics}
In this section, we focus on  two statistical properties of the wind power time series to summarise the future development of the basic characteristics of Danish wind power. These are the annual capacity factor (CF) and the variability of all Danish wind turbines (see definitions below). Both have a direct and important influence on technical as well as economical aspects of the power system.\\

The CF is defined in Eq.~\eqref{equation:CapacityFactor} below, and it is proportional to the total annual wind energy generation.
\begin{equation}\label{equation:CapacityFactor}
	\textup{CF} = \frac{\langle P \rangle_\textup{yr}}{P_\textup{rated}}
\end{equation}
where $\langle P \rangle_\textup{yr}$ is the hourly power generation averaged over one year, and $P_\textup{rated}$ is the total rated power.

\begin{figure}[tp!]
	\centering	

	\begin{subfigure}[b]{\columnwidth}	
	\includegraphics[width=\linewidth]{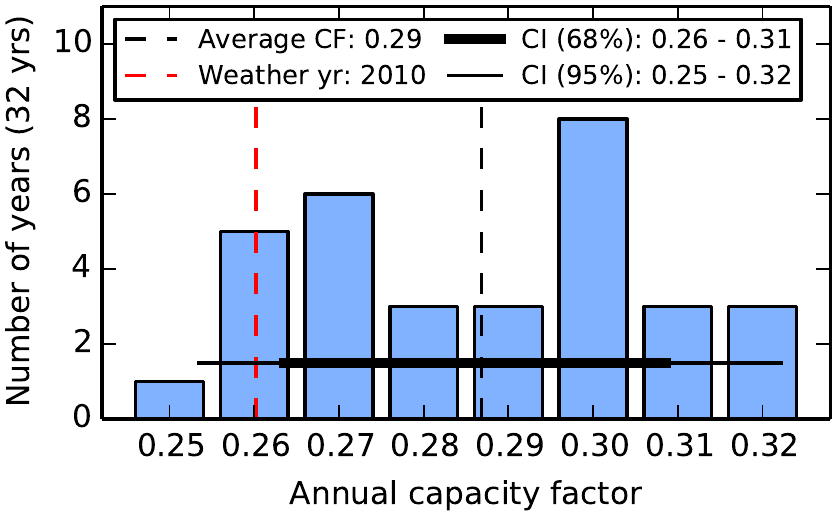}
		\caption{Model year: 2010}
	\end{subfigure}	
	\begin{subfigure}[b]{\columnwidth}
	\includegraphics[width=\linewidth]{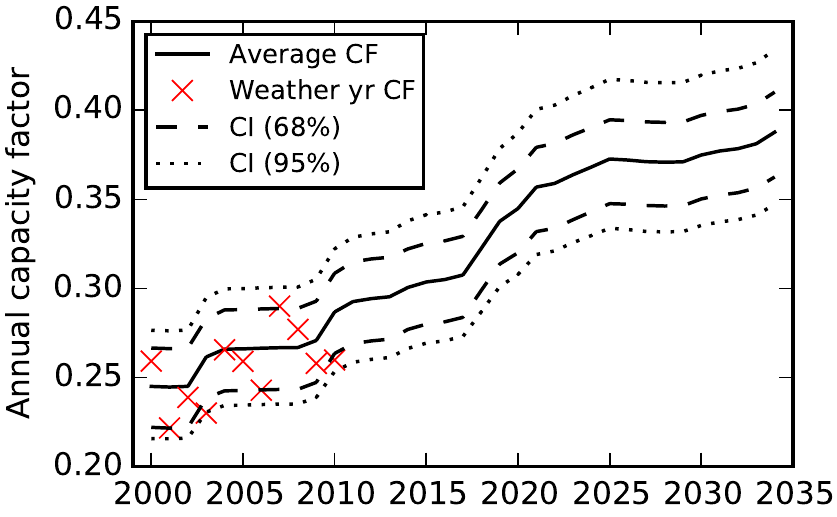}
		\caption{Model years: 2000 to 2035}
	\end{subfigure}	
	
	\caption{Variation in the annual capacity factor (CF). (a) Distribution of the annual capacity factor for all 32 weather years calculated for the turbine population of the model year 2010. The average CF of all weather years and the corresponding 68 and 95\% confidence intervals (CI) are indicated (black). The CF for the weather year 2010 is also shown (red). (b)  The average CF, and the 68 and 95\% confidence intervals are shown for all model years (black lines). The CF for the weather year corresponding to the model year is indicated (red crosses) for years until 2010. In all calculations the model year determines the population of wind turbines, and the power curve modification Eq.~\eqref{equation:smoothingFunction} is used with standard deviation $\sigma_0=2.29$~m/s, offset $\Delta\nu=1.27$~m/s, and availability $\eta=95\%$. }
	\label{figure:annualCF_2000to2035} 
\end{figure}

For the model year 2010, the distribution of {CF} for each of the 32 weather years in the {REatlas} is shown in Figure~\ref{figure:annualCF_2000to2035}a. The average CF of all weather years is 28.7\%, and the {CF}'s for the individual years are distributed nearly uniform and symmetrical around this value in the interval 25 to 32\%. This means that the total annual energy production of the 2010 population of Danish wind turbines will vary by more than $\pm10\%$ from year to year. A corresponding variation can be expected in related economical quantities such as the total annual feed-in tariff paid to support many of the turbines. As illustrated by Figure~\ref{figure:annualCF_2000to2035}b, the average CF increase roughly linearly from about 25\% in 2000 to about 39\% in 2035. The variation between individual weather years remains at about $\pm10\%$ of the average value. 

We note that this explains why the historical {CF} of the years 2000, 2005, and 2010 are all about equal despite the shift from small onshore to a combination of large onshore and offshore turbines in the period. The year 2000 was simply a good wind year, 2005 was average, and 2010 was poor with respect to wind power generation. A similar conclusion can be drawn from the national Danish wind energy index \cite{EMD:2013xy}, which is highly correlated with the results presented here.\\

The variability is a measure of the magnitude of hourly changes in wind power generation, and it can be related to the need for flexible up and down-regulation power capacity required to balance changes in wind power generation. In \cite{Holttinen:2008qd}, it is argued that the standard deviation of the annual distribution of hourly changes in wind power output can be used as a proxy for the variability. Here, we adopt a normalized variation of this measure of the variability with the definition:
\begin{equation}
	\textup{variability} = 100\%\times\frac{\textup{std}\left(\left\{P_{i+1}-P_i\right\}\right)}{P_\textup{rated}},
\end{equation}
where $\left\{P_{i+1}-P_i\right\}$ is the distribution of hourly differences in power generation, and $\textup{std}(X)$ denotes the standard deviation of $X$. According to \cite{Holttinen:2008qd}, this definition of variability should be multiplied by a factor of 3 to 6 times $P_\textup{rated}$ in order to estimate the corresponding need for up or down regulation. 

\begin{figure}[tp!]
	\centering	

	\begin{subfigure}[b]{\columnwidth}	
	\includegraphics[width=\linewidth]{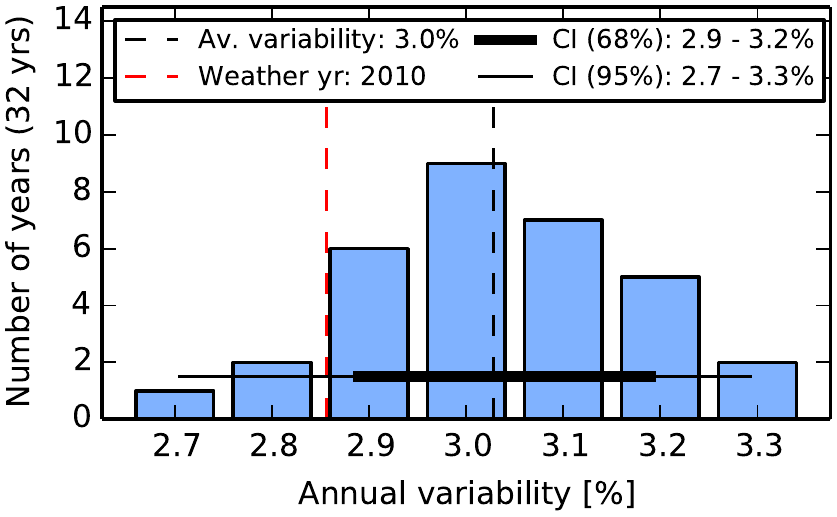}
		\caption{Model year: 2010}
	\end{subfigure}	
	\begin{subfigure}[b]{\columnwidth}
	\includegraphics[width=\linewidth]{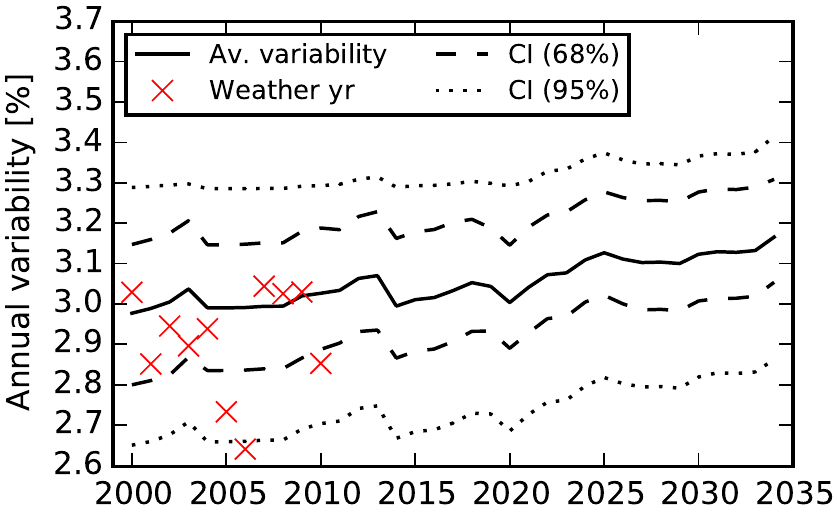}
		\caption{Model years: 2000 to 2035}
	\end{subfigure}	
	
	\caption{Variation in the annual variability. (a) Distribution of the annual variability for all 32 weather years calculated for the turbine population of the model year 2010. The average variability of all weather years and the corresponding 68 and 95\% confidence intervals (CI) are indicated (black). The variability for the weather year 2010 is also shown (red). (b)  The average variability, and the 68 and 95\% confidence intervals are shown for all model years (black lines). The variability for the weather year corresponding to the model year is indicated (red crosses) for years until 2010. In all calculations the model year determines the population of wind turbines, and the power curve modification Eq.~\eqref{equation:smoothingFunction} is used with standard deviation $\sigma_0=2.29$~m/s, offset $\Delta\nu=1.27$~m/s, and availability $\eta=95\%$. }
	\label{figure:annualVariability_2000to2035} 
\end{figure}

Between individual weather years, the variability generally deviates by about $\pm$10\% from the average value. This is illustrated in Figure~\ref{figure:annualVariability_2000to2035}a, where the distribution of the variability of the individual weather years is shown for the model year 2010. On average, the variability is 3.0\% in this case, and nearly all values fall in the 95\% confidence interval 2.7 to 3.3\%.

Unlike the CF, the normalized variability only changes slightly as the composition of the wind turbine population changes from 2000 to 2035. This is illustrated in Figure~\ref{figure:annualVariability_2000to2035}b, where the average variability increase by about 6 percent; from 3.0 in 2000 to 3.2\% in 2035. However, since up and down regulating reserves are often considered high value commodities, even small changes can result in important economic differences.

\section{Discussion: Wind power in 2020 energy system models}
\label{section:Discussion}
As mentioned in the introduction, fluctuations in the wind will clearly dominate many technical and economical aspects of the power system when the penetration of wind power in the Danish electricity system reach 50\% in 2020. Here, we use a handful of wind power time series from Danish 2020 scenarios, to discuss how differences in the characteristics of the time series may lead to important differences in the prediction of future characteristics of the power system. The motivation for this discussion is not to identify errors in existing analysis, but rather to point out that there is an increasing need for a consensus on how to select, characterise, and validate time series of wind power in the field of energy system modelling.

The time series have been collected from the sources listed below, and their annual capacity factor and and variability are shown in Table~\ref{table:ModelWindComparison2020}.
\begin{itemize}
	\item The {REatlas} 2020 and 2035 are described in Section~\ref{section:SelectedYearsWindPower} above. For each model year results for all 32 weather years are used.

	\item The {ISET} time series is a part of an 8-year-long European data set \cite{Bofinger:2008kx,Heide:2010fk}, where wind power modelling very similar to that described in Section~\ref{section:windConversion} was used. However, the onshore and offshore turbines were all assumed to have rated capacities of several {MW}, and the scenario is most similar to the 2035 model year of Section~\ref{section:SelectedYearsWindPower}.

	\item The {EnergyPLAN} model comes with publicly available input data. Here, the wind power time series of the study \cite{Mathiesen:2009uq} has been selected for comparison. In this case, a simple formula (see \cite{Lund:2010kx}) was used to modify historical wind power from 2001 to achieve a predetermined CF. 

	\item The RAMSES 6.12 model \cite{Danish-Energy-Agency:2010uo} is currently used by the Danish Energy Agency, and their wind power time series are based on historical production data from six different categories of wind turbines. Here, the time series have been combined according to the 2020 scenario of \cite{Energinet.dk:2013fk}.

	\item The final data set was obtained from the Danish Energy Association (DE). In this case, time series for four different categories of wind turbines are combined according to the 2020 scenario of \cite{Energinet.dk:2013fk}.
\end{itemize}

The {REatlas} and the {ISET} data sets both cover multiple years, and in both cases the CF and variability show similar relative variations from the average value. Thus, it is reasonable to assume that they include the typical annual variation in wind climate. The other data sets are limited to a single year of data, and the $\pm10$\% variation in both CF and variability between individual years is not captured directly in models based on this data. Since modern energy system models often contain a number of coupled and highly non-linear elements, it is not clear whether or not the effect of variation from year to year can be inferred statistically.

\begin{table}[b!]
	\begin{tabular}{l l l}
  		 & Capacity Factor & Variability \\
		\hline
		{REatlas} 2020 & 34.3\% (30.6;38.5) & 3.0\% (2.7;3.3)  \\
		{REatlas} 2035 & 39.1\% (35.0;43.7) & 3.2\% (2.9;3.4)  \\
		ISET & 40.9\% (36.5;44.8) & 3.6\% (3.3;3.9)  \\
		EnergyPLAN & 35.5\% & 3.7\%  \\
		RAMSES 6.12 & 36.1\% & 3.5\%  \\
		DE & 37.9\% & 2.5\%  \\
	\end{tabular}
	\caption{Annual capacity factor and variability for wind power time series used to model the Danish power system in 2020. For the {REatlas} and the {ISET} data sets, both mean value and the 95\% confidence interval is shown. All other time series cover a single year.}
	\label{table:ModelWindComparison2020}
\end{table}

The CF is about 35\% for the {REatlas}, {EnergyPLAN} and {RAMSES} 6.12 data sets, while it is respectively 37.9 and 40.9\% for the {DE} and the {ISET} data sets. This difference translates to about 10 to 15\% lower installed capacity for the the high {CF} data sets if the annual energy generation is kept constant. Consequently, hours with high wind power and low demand will typically occur less frequently in scenarios based on the high {CF} data, and when they occur they can be handled more easily. In a market simulation, a high {CF} can be expected to stabilise the hourly cost of electricity, because it indicates a less variable source of wind power.

As mentioned in Section~\ref{section:FutureCharacteristics}, the variability can be assumed to be proportional to the need for hourly up and down regulation reserves. For the {REatlas} 2020 data, the number varies between 2.5 and 3.3\%, the {ISET}, {EnergyPLAN}, and {RAMSES} 6.12 data sets all have significantly higher variability of about 3.6\%, and the {DE} data have a low variability of 2.5\%. This indicates that about 40\% more reserve capacity is required if the low variability data is exchanged with the high variability data.

In summary, the difference between current wind data sets is very likely to have significant effects on important technical and economical model predictions for 2020 and other future scenarios with high wind penetrations. In addition, it is unlikely that data sets that spans a single year can be used to produce robust results, since the annual variation in both CF and variability is relatively large.

\section{Conclusion}
\label{section:Conclusion}

Model time series of Danish wind power for each year in the period 1980 to 2035 have been calculated using a new state-of-the-art {REatlas} (renewable energy atlas). The time series are based on detailed representations of past and future Danish wind turbines, and a calibration and validation against historical data have been carried out. For each model year, data from 32 weather years are available.

The time series have been used to show how the average annual capacity from Danish wind turbines increase from about 25\% in 2000 to almost 40\% in 2035 as old onshore turbines are gradually replaced by new large onshore and offshore turbines. It is also shown that the annual CF varies by $\pm10$ for both past and future years. The annual need for up and down regulation reserves caused by hourly variations in the Danish wind power generation varies by a similar percentage, but it does not increase significantly in the future per MW wind installed.

Finally, a number of time series used in models of Danish 2020 scenarios with 50\% penetrations of wind power in the electricity system have been compared. The motivation is to highlight the need for a consensus on how to select, characterise, and validate such data in the field of energy system modelling. With this in mind, the calibrated Danish model time series of this paper are all freely available for download \cite{Andresen:2013uq}.

\bigskip
\noindent {\bf Acknowledgements:} The authors would like to thank the Danish Energy Agency, the Danish Energy Association, and the Sustainable Energy Planning Research Group at Aalborg University for help with and access to their Danish wind power time series for 2020.

This work was financially supported by Dansk Energi, DONG Energy and The Danish National Advanced Technology Foundation (j.nr. 140-2012-5). 

\bibliography{references} 

\appendix
\section{The {RE}atlas software and hardware implementation}\label{appendix:conversionSetup}

In this section, the {REatlas} conversion process is described in detail and the choice of hardware is motivated. The primary measure of success used for the implementation is to reduce the time it takes to perform repeated conversions for the same subset of grid points. 

The overall process of converting weather data from the CFSR data set (see Section~\ref{section:weatherData}) to wind or solar power generation is divided into two main steps: First, the CFSR data is decompressed and a subset containing data for the region and time period of interest is organised in an uncompressed array. The implementation of this step is specific to the CFSR data format, and it is relatively time consuming. In the second step, weather data is converted to wind or solar power generation. Here, the uncompressed array is used as input data, which means that it is independent of the original data source. The second step is very fast when compared to the first step. For this reason it is ideal for repeated conversions with different types of wind or solar technologies.

Below, the motivation for splitting the conversion process into two steps and for implementing the {REatlas} on high-end hardware is treated first. This description is followed by a more detailed treatment of the most important bottlenecks in the initial preparation step and in the conversion step. We do not provide details of the {RE}atlas source code which is mainly written in the high-level language {Python} with selected subroutines written in {C} to ensure efficient memory management. In our setup, the latter gives a speed increase of about a factor of 10.

\subsection*{General motivation}

When implemented as directly as possible, the complete conversion of the entire data set takes a little less than a week to complete on a standard 8 threaded desktop PC of 2012. If one were to save the result back to {GRIB} files, this step is estimated to take about as long. In a test run, it was observed that most of the time was spent on decompression of the original {GRIB} files (and compression of the resulting time series). This is because the conversion algorithm is, arithmetically speaking, extremely simple compared to data compression algorithms. 

The simplest way to overcome this obstacle is to store the weather data uncompressed. But the size of the entire 32-year-long global data set would increase from about 700~GB compressed to about 10~TB uncompressed. However, if only regions of interest are stored uncompressed, the data size can be brought to a more manageable size. As an example, a European data set with 280,320 grid points and 280,512~hours (32~years) takes up about 88~GB when stored as uncompressed single precision floating point numbers. On a standard desktop PC, decompression of the CFSR data used in the {REatlas} takes about one week, and on the high-end computer currently used for the {REatlas} (see Section~\ref{section:REatlasImplementation}) it takes about 20~h. It was decided that this was much too long for repeated conversions, and as a consequence the process was split into an initial step where raw data is prepared in a standardised uncompressed format and a second step where the actual conversion from weather data to power generation happens.

An added benefit is that only a subset of the global data set needs to be processed during the conversion step. When working directly with the {GRIB} files, each global time slice is compressed separately, and data for the entire world must be decompressed even if only data from a single location is converted to power generation. In addition, the conversion step becomes independent of the original data source, which can then be replaced more easily.

A test conversion of the European data set showed a speed up of a factor of 140, from one week to a little over one hour on a standard desktop PC when an uncompressed array of weather data was used in place of the original CFSR data files. This time corresponds quite well to the time it takes to move 88~GB off a hard drive with typical bus speed of about 50~MB/s. Based on this test, it was decided to move the data to a memory mapped array before the conversion to minimise this bottleneck. Typical memory peak transfer rates are currently in excess of 10~GB/s, which, in principle, allows for an additional conversion speed increase by a factor of more than 200.

Standard PC's does not currently have several hundred GB memory capacity, so to realise the conversion speed gain from a memory mapped uncompressed array of input data, the {REatlas} is implemented on a high-end computer as described in Section~\ref{section:REatlasImplementation}. On this computer conversion speeds for the European data set are down to 45~s for wind power and less than 2~minutes for solar power. In this case, the dominating time limiting bottleneck is processing and not memory speed.

The memory mapped array is realised by moving the data to a so called {RAM} disk storage (RD storage), which the computer treats as a virtual hard drive. Among other things, this gives the advantage that the software can easily be run on computers with less memory. As an example a solid state drive could be used as an alternative to RAM. Working at a bus speed of 3~Gbit/s it would transfer 88~GB in about 6~minutes.

\subsection*{Step 1: Initial preparation of weather data}

\begin{figure*}[t]
	\centering	
	\includegraphics[width=0.8\linewidth]{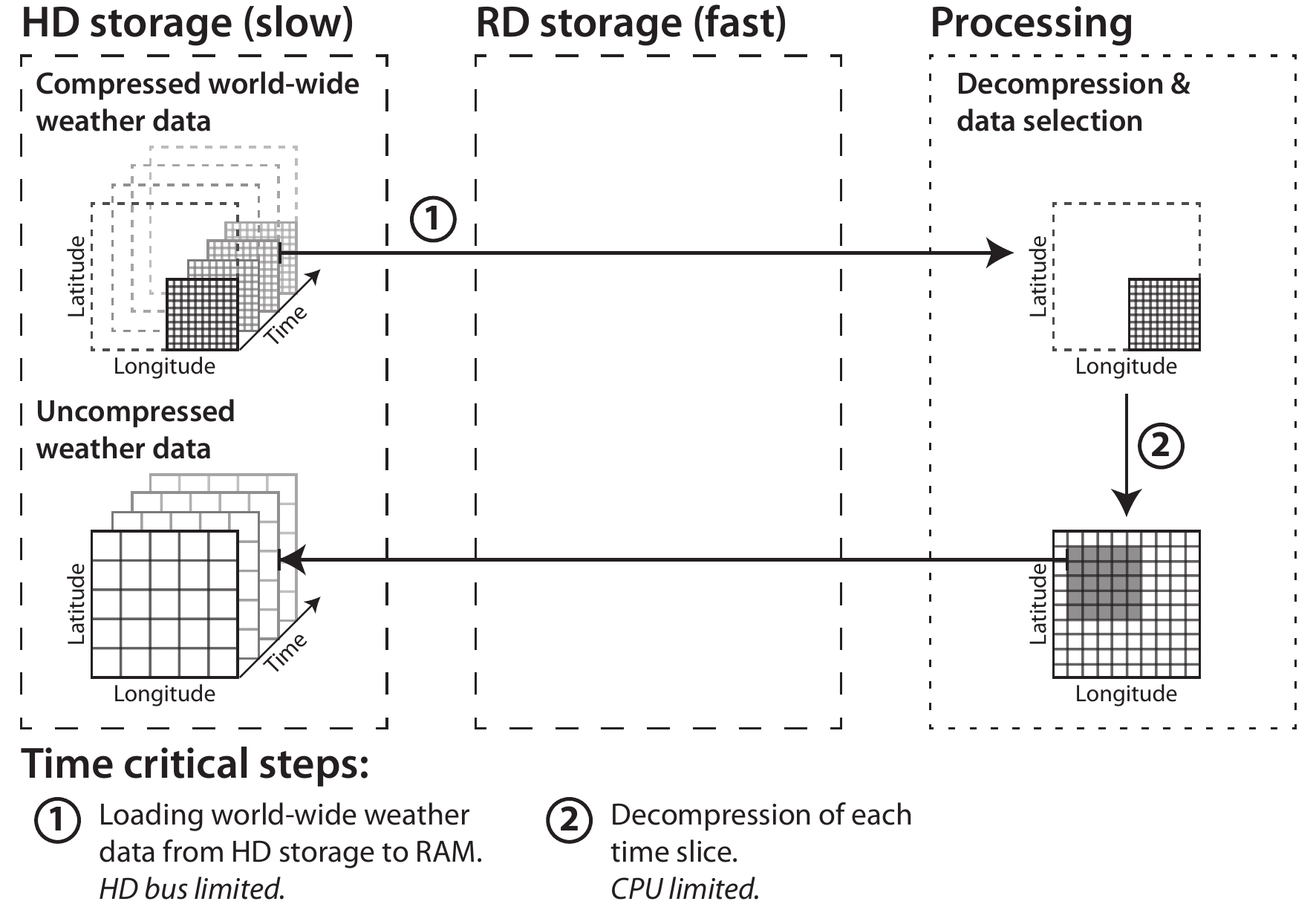}	
	\caption{Schematic representation of \emph{Step 1: Initial preparation of weather data}. Note that RD storage is bypassed in this step.}
	\label{figure:REatlasCutoutStep} 
\end{figure*}

In the first step of the conversion process a region of interest is selected. The entire set of global weather data is then decompressed one time slice at a time, and a new time slice containing only the data for the selected subset of the world is stored in a binary array. This process is illustrated schematically in Figure~\ref{figure:REatlasCutoutStep}, where the two most time critical bottlenecks are indicated. Note that RD storage is bypassed in this step.

The first bottleneck marked \circled{1}, is the transfer of the uncompressed {GRIB} files containing the original {CFSR} data from storage to the processing unit. This process is limited by the HD bus speed, or in our setup by a combination of the network area storage (SAN) speed and the network connection. The {REatlas} is currently based on about 700~GB of data from the {CFSR} data set (see Section~\ref{section:weatherData}). In the current setup, we are able to transfer data from SAN to memory with about 200~MB/s resulting in a total transfer time of less than one hour. At a typical HD transfer rate of 50~MB/s, the transfer takes a few hours. 
 
The second bottleneck marked \circled{2} is decompression of the {GRIB} files. The {GRIB} files contain global hourly time slices (also called records), one for each data field used in the wind or solar conversions. An example could be wind speed at 10~m height. The time slices are compressed separately using the {JPEG2000} algorithmn, which means that the entire time slice must be decompressed in order to select a specific subset of the grid points. In the current setup, the total decompression time is about 20~h. On a standard PC, the corresponding time is one week. This is far longer than bottleneck \circled{1}, showing that bottleneck \circled{2} clearly dominates.

\subsection*{Step 2: Conversion to wind or solar power generation}

\begin{figure*}[t]
	\centering	
	\includegraphics[width=0.8\linewidth]{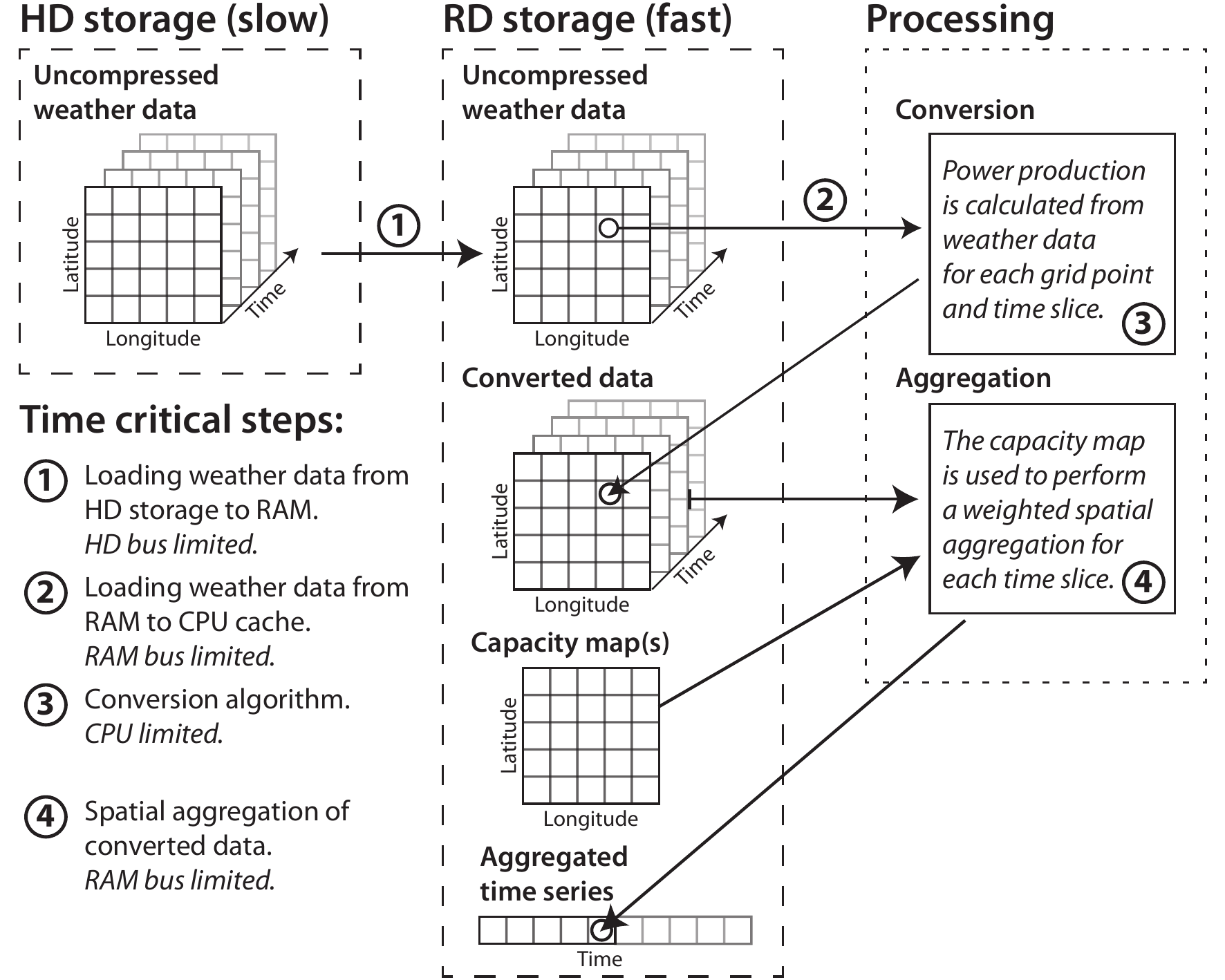}	
	\caption{Schematic representation of \emph{Step 2: Conversion to wind or solar power generation}. The uncompressed weather data may contain multiple values per grid point, e.g. wind speed and roughness.}
	\label{figure:REatlasConversionStep} 
\end{figure*}

In the second step of the conversion process, the uncompressed cutout is first moved from slow HD storage into the much faster RD storage. A memory mapped array is then allocated for the output data, and then wind or solar power production is calculated for each point in time and space. Finally, the calculated power generation is aggregated spatially for each time slice using a fixed capacity map, supplied by the user. If the same data set is converted multiple times, allocation of the in and output arrays in RD storage and the initial transfer from HD storage is only done once.

The process is illustrated schematically in Figure~\ref{figure:REatlasConversionStep}, where the four most time critical bottlenecks are indicated. In the first of these, marked \circled{1}, the uncompressed weather data is moved from HD storage to RD storage. This process is limited by the HD bus speed and the size of the cutout. In the current setup, data is stored on network area storage (SAN) connected with Gigabit Ethernet. For the 88~GB European data set, the transfer is typically completed in about 10~min. On a standard desktop PC, the transfer would take about half an hour at a typical speed of 50~MB/s.

The second bottleneck marked \circled{2} is reading the raw data from RD storage to the CPU cache. The maximum peak transfer rate of the current setup is 102.4~GB/s, and at this speed the entire 512~GB memory block can be processed in 5~s. For the European data set, we observe a processing time of about 6~s when performing an identity operation in place of the wind conversion. This corresponding to a transfer rate of 8~GB/s, which includes reading about 22~GB of wind speed data and writing the result back to memory. The discrepancy between the maximum peak transfer rate and the measured rate is not uncommon for multichannel memory.

The bottleneck marked \circled{3}, is the conversion algorithm. For both wind and solar conversions, the algorithm is simple in the sense that only relatively few floating point operations (flops) are required. A single modern CPU can perform on the order of 100 to 1000 flops in the time it takes to transfer a number from memory to the CPU, and the trade off between the bottlenecks \circled{2} and \circled{3} is determined by the ratio between memory bus speed and the total number and speed of the CPU's. In our setup, we observe conversion times where the processing speed is the limiting factor. Thus, bottleneck \circled{3} dominates bottleneck \circled{2} in this case.

In a setup with additional processing power, it could make sense to use LZO compression to increase the rate at which data is delivered to the processors. The LZO compression algorithm is designed to be extremely fast at decompression. Therefore, if you compress data with LZO, you will be able to move it faster without spending significant amounts of time decompressing it afterwards.

The last bottleneck \circled{4} indicated on Figure~\ref{figure:REatlasConversionStep} is spatial aggregation of the converted data. Here, the weighted sum of each time slice is calculated. This is a very simple algebraic operation, and the limiting factor is the memory bus speed. In principle, aggregation can be performed as part of the conversion process, thus eliminating this bottleneck. However, in the current implementation it was decided to separate the two processes, since aggregation is not always needed and combining them complicates memory management.

Finally, the accuracy of both in and output data is an important factor in both data size and processing speed. As an example, using single precision floating point numbers instead of double precision numbers gives an overall speed gain of about a factor of 2 and halves the data size. However, some care should be taken when the precision is reduced if the numbers vary by several orders of magnitude. Currently, single precision floats is used for standard operation in the {REatlas}.

\end{document}